\documentclass[conference,letterpaper]{IEEEtran}
\IEEEoverridecommandlockouts 
\usepackage[utf8]{inputenc} 
\usepackage[T1]{fontenc}
\usepackage{url}
\usepackage{ifthen}
\usepackage{cite}
\usepackage[cmex10]{amsmath} 
\usepackage{amssymb}
\usepackage{amsfonts}
\usepackage{graphicx}
\usepackage{overpic}
\newtheorem{thm}{Theorem}
\newtheorem{lem}{Lemma}
\newtheorem{cor}{Corollary}
\newcommand{\ra}{\rightarrow}

\newcommand{\teq}{\triangleq}

\newcommand{\cA}{\mathcal{A}}
\newcommand{\cB}{\mathcal{B}}

\newcommand{\cE}{\mathcal{E}}
\newcommand{\cF}{\mathcal{F}}
\newcommand{\cG}{\mathcal{G}}

\newcommand{\cN}{\mathcal{N}}

\newcommand{\cP}{\mathcal{P}}

\newcommand{\cS}{\mathcal{S}}

\newcommand{\cV}{\mathcal{V}}

\newcommand{\argmax}{\mathop\mathrm{argmax}\limits}

\newcommand{\bfC}{\mathbf{C}}
\newcommand{\bfD}{\mathbf{D}}

\def\eqo#1{\overset{\mathrm{#1}}{=}}
\def\geqo#1{\overset{\mathrm{#1}}{\geq}}

\begin{document}
\title{On the Distance Between the Rumor Source\\ and Its Optimal Estimate in a Regular Tree\thanks{An earlier version was presented at SITA2014 \cite{matsuta2014pdd}. In this paper, we improved notations, added Corollary \ref{cor:withinD}, revised proofs, and corrected the bound of Theorem \ref{thm:lower upper any delta} and many errors.}} 


\author{
  \IEEEauthorblockN{Tetsunao Matsuta\IEEEauthorrefmark{1} and Tomohiko Uyematsu\IEEEauthorrefmark{2}}
  \IEEEauthorblockA{Department of Information and Communications Engineering, Tokyo Institute of Technology
  }
  \IEEEauthorblockA{Email: tetsu@ict.e.titech.ac.jp\IEEEauthorrefmark{1}, uyematsu@ieee.org\IEEEauthorrefmark{2}}
}



\maketitle

\begin{abstract}
    This paper addresses the rumor source identification problem, where the goal is to find the origin node of a rumor in a network among a given set of nodes with the rumor. In this paper, we focus on a network represented by a regular tree which does not have any cycle and in which all nodes have the same number of edges connected to a node. For this network, we clarify that, with quite high probability, the origin node is within the distance ``$3$'' from the node selected by the optimal estimator, where the distance is the number of edges of the unique path connecting two nodes. This is clarified by the probability distribution of the distance between the origin and the selected node.
\end{abstract}


\section{Introduction}
In social networks, a rumor spreads like an infectious disease. In fact, it can be modeled as an infectious disease \cite{bailey1975mathematical,shah2011rumors}. The most common theme of studies about a rumor (or infectious disease) is to analyze mechanisms of a spreading behavior of a rumor in a given network \cite{pastor2001epidemic,may2001infection}.

Unlike this type of studies, we address the {\it rumor source identification problem} introduced by Shah and Zaman \cite{shah2011rumors}. The goal of this problem is to find the origin node of a rumor (rumor source) in a network among a given set of nodes with the rumor. If the rumor source can be detected, it is available to find a weak node which spreads a computer virus, to give ranking to websites for a search engine, etc. For this problem, Shah and Zaman \cite{shah2011rumors} introduced the optimal estimator and analyzed the correct detection probability of it for some types of networks. This probability asymptotically goes to one for a very special network called geometric tree (see \cite[Sec.\ IV.D]{shah2011rumors}). However, they analytically or experimentally showed that the probability is asymptotically not high or goes to zero for many other networks such as regular trees, small-world networks, and scale-free networks, where a regular tree is a network which does not have any cycle and in which all nodes have the same degree, i.e, the number of edges connected to a node.

Although the optimal estimator may not find the rumor source, it actually selects a node near the rumor source. This fact is known experimentally (cf.\ \cite[Sect.\ V.B]{shah2011rumors} and \cite[Sect.\ 8]{7740106}) and is not known analytically to the best of our knowledge. In this paper, we focus on this fact and clarify it analytically. Especially, we focus on regular trees and clarify that, with quite high probability, the rumor source is within the distance ``$3$'' from the node selected by the optimal estimator, where the distance is the number of edges of the unique path connecting two nodes. This is clarified by the probability distribution of the distance between the rumor source and the selected node.

\section{Rumor Source Identification Problem}
In this section, we introduce the rumor source identification problem and show some known results of this problem.

Let $\cG$ be an undirected and connected graph. Let $\cV(\cG)$ denote the set of nodes and $\cE(\cG)$ denote the set of edges of the graph $\cG$. We denote the edge connecting two nodes $i,j \in \cV(\cG)$ by the set of nodes $\{i, j\} \in \cE(\cG)$. In this paper, we consider the case where $\cG$ is a regular tree, that is, the graph does not have any cycle, and all nodes have the same degree\footnote[2]{The line graph ($\delta=2$) is not concerned in this paper because this case is somewhat difficult to treat in a unified manner. However, essential argument for this case is the same as the case where $\delta\geq 3$.} $\delta\geq 3$. We assume that the number of nodes is countably infinite in order to avoid boundary effects.

A rumor spreads in a given regular tree $\cG$. Initially, the only one node $v_1\in\cV(\cG)$ (the rumor source) possesses a rumor. The node possessing the rumor infects it to connected adjacent nodes, and these nodes keep it forever. For $\{i, j\} \in \cE(\cG)$, let $\tau_{ij}\in\mathbb{R}$ be a real-valued random variable (RV) that represents the rumor spreading time from the node $i$ to the node $j$ after $i$ gets the rumor. In this model, spreading times $\{\tau_{ij}: \{i, j\} \in \cE(\cG)\}$ are independent and drawn according to the exponential distribution with the unit mean. Thus, the cumulative distribution function $F$ of $\tau_{ij}$ is represented as $F(x) = 1 - e^{-x}$ if $x\geq 0$, and $F(x) = 0$ if $x\leq 0$. This spreading model is sometimes called the susceptible-infected (SI) model \cite{shah2011rumors}.

Suppose that we observe a network consisted of $n$ infected nodes in the graph $\cG$ at some time. Since the rumor spreads to the connected adjacent nodes, this network is a connected subgraph of $\cG$. We denote the RV of this network by $G_n$ and its realization as $\cG_n$. We only know an observed network and do not know the realization of spreading times on edges. Then, the goal of the rumor source identification problem is to find the rumor source $v_1$ among $\cV(\cG_n)$ given $\cG_n$.

For this problem, the optimal estimator is the maximum likelihood (ML) estimator $\varphi_{\mathrm{ML}}(\cG_n)$ (cf.\ \cite{shah2011rumors}) defined as
\begin{align*}
    \varphi_{\mathrm{ML}}(\cG_n) \teq \argmax_{v\in\cV(\cG_n)} \Pr\{\cG_n|v\},
\end{align*}
where ties broken uniformly at random and $\Pr\{\cG_n|v\}$ is the probability observing $\cG_n$ under the SI model assuming $v$ is the rumor source. For this optimal estimator, let $\bfC_{n}$ be the correct detection probability when a graph of $n$ infected nodes is observed, i.e., $\bfC_{n} = \Pr\{\varphi_{\mathrm{ML}}(G_n) = v_1\}$. Shah and Zaman \cite{shah2016finding} showed the asymptotic behavior of  $\bfC_{n}$ as the next theorem.

\begin{thm}[{\cite[Theorem 3.1]{shah2016finding}}]
    \label{thm:shah and zaman}
    For a regular tree with degree $\delta\geq 3$, it holds that
    \begin{align}
        \lim_{n\ra\infty}\bfC_{n} = \delta I_{1/2}\left(\frac{1}{\delta-2},\frac{\delta-1}{\delta-2}\right)
        -(\delta-1),
        \label{equ:CorrectProb}
    \end{align}
    where $I_{x}(a,b)$ is the regularized incomplete beta function defined as $I_{x}(a,b) \teq \frac{\Gamma(a+b)}{\Gamma(a)\Gamma(b)}\int_{0}^{x}t^{a-1}(1-t)^{b-1} dt,$ and $\Gamma(\cdot)$ is the Gamma function.
\end{thm}

According to this theorem, when $\delta = 3$, $\lim_{n\ra\infty}\bfC_{n} = 0.25$. Moreover, it rapidly converges to $1 - \ln(2) \approx 0.307$ as $\delta$ goes to infinity (cf.\ \cite[Corollary 1 and Figure 3]{shah2016finding}). This means that, unfortunately, the correct detection probability is \textit{not} very high for regular trees.

\section{Main Results}
In this section, we show that the ML estimator can select a node near the rumor source with high probability.

To this end, we clarify the probability distribution of the distance $d\ (\geq 1)$ between the rumor source and the node selected by the ML estimator. We denote this probability by $\bfD_{n}(d)$ and define it as
\begin{align}
    \bfD_{n}(d) \teq \Pr\{ d_{\cG}(\hat V_{n}, v_1) = d \},
    \label{equ:distance prob}
\end{align}
where $\hat V_{n} = \varphi_{\mathrm{ML}}(G_n)$ and $d_{\cG}(v, w)$ denotes the distance between nodes $v$ and $w$ in the graph $\cG$. Note that $\bfD_{n}(0) = \bfC_{n}$.

When $\delta = 3$, we can clarify a closed-form expression of the asymptotic behavior of $\bfD_{n}(d)$ as the next theorem.
\begin{thm}
    \label{thm:asymptotic delta=3}
    Let $\delta=3$. Then, for any $d\geq 1$, we have
    \begin{align*}
        \lim_{n\ra\infty}\bfD_{n}(d) = f(d),
    \end{align*}
    where
    \begin{align*}
        f(d) &= 3\cdot 2^{d-1} (-1)^{d} \\
        &\quad \times \left(\frac{1}{4} + \sum_{l=1}^{d} (-1)^l\left( \frac{(\ln 2)^l}{l!} -2 + \sum_{m=0}^l \frac{(\ln 2)^m}{m!} \right) \right).
    \end{align*}
\end{thm}

We denote the rising factorial $x(x+1)(x+2)\cdots(x+k-1)$ by $x^{\overline k}$.
The next theorem gives tight upper and lower bounds of $\lim_{n\ra\infty} \bfD_n(d)$ for more general degrees.
\begin{thm}
    \label{thm:lower upper any delta}
    For any $\delta\geq 3$, $d\geq 1$, and $m \geq d + 1$, we have
    \begin{align*}
        0\leq \lim_{n\ra\infty} \bfD_n(d) - g(\delta,d,m) \leq \epsilon_m,
    \end{align*} 
    where $\epsilon_m = e^2 (8 + 5 m + m^2) 2^{- m + 3}$,
    \begin{align*}
        g(\delta,d,m)
        &= \delta(\delta-1)^{d-1} \sum_{k=d+1}^{m} p_{1}(\delta, d, k) p_{2}(\delta, k),\\
        p_{1}(\delta, d, k) &= \frac{2}{(\delta - 2)^{d}}
        \frac{(\frac{1}{\delta - 2} )^{\overline{k-1}}}
        {(\frac{2}{\delta - 2})^{\overline{k}}}
        \zeta_{k-2}^{d-1}\left(\frac{1}{\delta-2}\right),\\
        p_{2}(\delta,k) &= I_{1/2}\left(k-1+\frac{1}{\delta-2}, \frac{\delta-1}{\delta-2}\right)\\
        &\quad - (\delta-1)I_{1/2}\left(k-1+\frac{\delta-1}{\delta-2},\frac{1}{\delta-2}\right),
    \end{align*}
    $\zeta_{k}^{d}(x) = \sum_{1\leq j_1 < j_2<\cdots<j_{d} \leq k} \left(\prod_{i=1}^{d} \frac{1}{j_i+x} \right)$, and $\zeta_{k}^{0}(x) = 1$ for any $k \geq 0$.
\end{thm}

$\zeta_{k}^{d}(x)$ is a partial sum of the multiple Hurwitz zeta function (cf.\ e.g.\ \cite{murty2006multiple})
or the \textit{shifted} multiple harmonic sums (cf.\ e.g.\ \cite{hoffman1992multiple}). We note that the difference of bounds (i.e., $\epsilon_{m}$) does not depend on degrees.

\begin{figure}[t!]
    \centering
    \begin{overpic}[width=80mm]{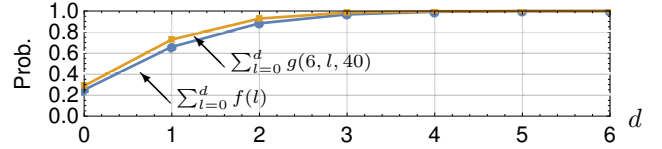}
        \put(102, 3){$d$}
        \put(35, 13){\vector(-1,1){5} \scriptsize $\sum_{l = 0}^{d} g(6, l, 40)$}
        \put(26, 7){\vector(-1,1){5} \scriptsize $\sum_{l = 0}^{d} f(l)$}
    \end{overpic}
    \caption{$\sum_{l = 0}^{d} f(l)$ and $\sum_{l = 0}^{d} g(6, l, 40)$. $\sum_{l = 0}^{3} f(l) \approx 0.968$. $\sum_{l = 0}^{3} g(6, l, 40) \approx 0.985$.}
    \label{fig:prob}
\end{figure}

These theorems imply that the ML estimator can select a node near the rumor source with high probability. This is clear from the next corollary and its numerical results (Fig.\ \ref{fig:prob}).
\begin{cor}
    \label{cor:withinD}
    Let $\delta = 3$. Then, for any $d \geq 0$, we have
    \begin{align*}
        \lim_{n \ra \infty} \Pr\{ d_{\cG}(\hat V_{n}, v_1) \leq d \} = \sum_{l = 0}^{d} f(l).
    \end{align*}
    More generally, for any $\delta\geq 3$, $d\geq 0$, and $m \geq d + 1$, we have
    \begin{align*}
        0 \leq \lim_{n \ra \infty} \Pr\{ d_{\cG}(\hat V_{n}, v_1) \leq d \} - \sum_{l = 0}^{d} g(\delta, l, m) \leq d \cdot \epsilon_m.
    \end{align*}
    Here, $f(0)$ and $g(\delta, 0, m)$ denote the right-hand side of \eqref{equ:CorrectProb}.
\end{cor}
\begin{IEEEproof}
    By noticing that $\Pr\{ d_{\cG}(\hat V_{n}, v_1) \leq d \} = \sum_{l = 0}^{d} \bfD_{n}(l)$, the corollary is immediately obtained by Theorems \ref{thm:shah and zaman}-\ref{thm:lower upper any delta}.
\end{IEEEproof}

\begin{figure}[t!]
    \centering
    \begin{overpic}[width=80mm]{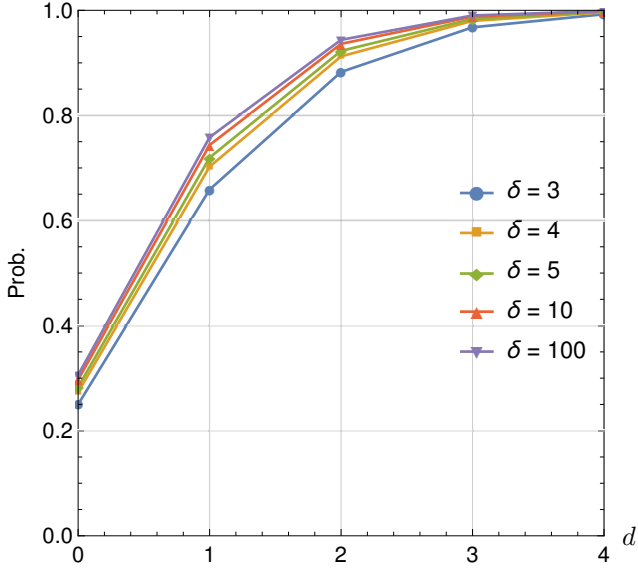}
        \put(102, 3){$d$}
    \end{overpic}
    \caption{$\sum_{l = 0}^{d} f(l)$ (i.e., the case where $\delta = 3$) and $\sum_{l = 0}^{d} g(\delta, l, 40)$ (i.e., the case where $\delta \geq 4$).}
    \label{fig:prob_others}
\end{figure}

Since $\epsilon_{40} \approx 10^{-7}$, Fig.\ \ref{fig:prob} gives almost exact numerical results of $\lim_{n \ra \infty} \Pr\{ d_{\cG}(\hat V_{n}, v_1) \leq d \}$. We note that numerical results for other degrees $\delta$ are almost the same (see Fig.\ \ref{fig:prob_others}). Thus, these results show that the rumor source is within the distance $3$ from the node selected by the ML estimator with quite high probability. We note that Khim and Loh \cite[Corollary 2]{7740106} gave another lower bound of $\lim_{n \ra \infty} \Pr\{ d_{\cG}(\hat V_{n}, v_1) \leq d \}$. However, it is quite looser than our bound and is zero at least values of parameters $d$ and $\delta$ are within the rage in Fig.\ \ref{fig:prob} and Fig.\ \ref{fig:prob_others}.

\section{Proofs of Theorems}
In this section, we prove our main theorems. We will denote $n$-length sequences of RVs $(X_1,X_2,\cdots,X_n)$ and its realizations $(x_1,x_2,\cdots,x_n)$ by $X^n$ and $x^n$, respectively. For the sake of brevity, we denote $\cV(\cG)$ by $\cV$ and $\cV(G_n)$ by $\cV_n$.

For any node $v \in \cV$ in a regular tree with degree $\delta\geq 3$, there are $\delta$ neighbors. Thus, there are $\delta$ subtrees rooted at these $\delta$ neighbors with the parent node $v$. In other words, the regular tree is divided into these $\delta$ subtrees and the node $v$. Let $X_j(v)$ be the number of infected nodes in the $j$th subtree among those subtrees ($j=1,2\cdots,\delta$). When $v$ is not the rumor source, let $\delta$th subtree contain the rumor source $v_1$. Note that, if $v$ is an infected node, we have $\sum_{j=1}^\delta X_j(v)=n-1$. The next lemma is a key lemma to prove our main theorems.
\begin{lem}
    \label{lem:correct prob with cond.}
    For a node $v \in \cV$, let $\overline X(v) = \max_{1\leq j \leq \delta}\{X_j(v)\}$. Then, we have
    \begin{align*}
        \Pr\{\hat V_{n}=v| v \in \cV_n, \overline X(v) < n/2 \}&= 1,\\
        \Pr\{\hat V_{n}=v| v \in \cV_n, \overline X(v) = n/2 \}&= 1/2,\\
        \Pr\{\hat V_{n}=v| v \in \cV_n, \overline X(v) > n/2 \}&= 0.
    \end{align*}
\end{lem}

Since this lemma can be obtained by \cite[Proposition 1]{dong2013rootingArXiv4} (see also {\cite[Lemma 6]{dong2013rootingArXiv4}}), we prove this in Appendix \ref{app:rumor_centrality}.

We denote the set of nodes with distance $d\ (\geq 1)$ from the rumor source by $\cV^{(d)}$. Note that the number of elements of $\cV^{(d)}$ is $\delta (\delta-1)^{d-1}$. Then, $\bfD_{n}(d)$ can be represented as
\begin{align}
    \bfD_{n}(d)
    &= \Pr\{ d_{\cG}(\hat V_{n}, v_1) = d \}\notag\\
    &= \Pr\{\hat V_{n} \in \cV^{(d)} \}\notag\\
    &= \sum_{v^{(d)} \in \cV^{(d)}}\Pr\{\hat V_{n} = v^{(d)}\}\notag\\    
    &= \sum_{v^{(d)} \in \cV^{(d)}} \Pr\{v^{(d)}\in\cV_n, \hat V_{n} = v^{(d)} \}\notag\\
    &= \sum_{v^{(d)} \in \cV^{(d)}} \left( \Pr\{ v^{(d)}\in\cV_n, \hat V_{n} = v^{(d)}, \overline X(v^{(d)})<n/2 \} \right. \notag\\
    &\quad +\Pr\{v^{(d)}\in\cV_n, \hat V_{n} = v^{(d)}, \overline X(v^{(d)})=n/2 \}\notag\\
    &\quad \left. +\Pr\{v^{(d)}\in\cV_n, \hat V_{n} = v^{(d)}, \overline X(v^{(d)})>n/2 \}\right)\notag\\
    &= \sum_{v^{(d)} \in \cV^{(d)}} \left(\Pr\{v^{(d)}\in\cV_n, \overline X(v^{(d)})<n/2\}\right.\notag\\
    &\quad \left. +1/2\Pr\{v^{(d)}\in\cV_n,
        \overline X(v^{(d)})=n/2\}\right),
    \label{equ:distance prob with lemma}
\end{align}
where the last equality comes from Lemma \ref{lem:correct prob with cond.}.

On the other hand, let $\{V_i\}_{i=1}^{\infty}$ be the sequence of RVs each representing $i$th infected node, where $V_1=v_1$ with probability 1. Then, we have $\cV_n=\{V_1,V_2,\cdots,V_n\}$. This implies that the event $\{v^{(d)} \in \cV_n\}$ is equal to the event $\cup_{k=d+1}^{n}\{V_k = v^{(d)}\}$. Hence, we have
\begin{align}
    &\Pr\{v^{(d)}\in\cV_n, \overline X(v^{(d)})<n/2 \}\notag\\
    &= \Pr\{\cup_{k=d+1}^{n}\{V_k=v^{(d)}\},
    \overline X(v^{(d)})<n/2\}\notag\\
    &= \Pr\{\cup_{k=d+1}^{n}\{V_k=v^{(d)},
    \overline X(v^{(d)})<n/2\}\}\notag\\
    &= \sum_{k=d+1}^{n}\Pr\{V_k=v^{(d)},
    \overline X(v^{(d)})<n/2\}\notag\\
    &= \sum_{k=d+1}^{\lceil n/2 \rceil}\Pr\{V_k=v^{(d)},
    \overline X(v^{(d)})<n/2\}\notag\\
    &= \sum_{k=d+1}^{\lceil n/2 \rceil} \sum_{\substack{x^\delta:\sum_{j=1}^\delta x_j=n-1,\\ \max_{1\leq j \leq \delta}\{x_j\} < n/2}}\hspace{-5mm} \Pr\{V_k=v^{(d)}, X^\delta(v^{(d)}) = x^\delta\},
    \label{equ:correct prob < n/2}
\end{align}
where $X^\delta(v)=(X_1(v),X_2(v),\cdots,X_\delta(v))$. We also have
\begin{align*}
    &\Pr\{v^{(d)}\in\cV_n,
    \overline X(v^{(d)})=n/2\}\\
    &= \sum_{k=d+1}^{\lfloor n/2 \rfloor + 1} \sum_{\substack{x^\delta:\sum_{j=1}^\delta x_j=n-1,\\ \max_{1\leq j \leq \delta}\{x_j\} = n/2}}\Pr\{V_k=v^{(d)}, X^\delta(v^{(d)}) = x^\delta\}.
\end{align*}

Thus, we need to obtain closed-form expressions of $\Pr\{V_k=v^{(d)}\}$ and $\Pr\{X^\delta(v^{(d)}) = x^\delta| V_k=v^{(d)}\}$.
\subsection{Closed-Form Expression of $\Pr\{V_k=v^{(d)}\}$}
Let $\cN(v)$ be the set of neighboring nodes of $v$ in the graph $\cG$.
Suppose that the set $\hat \cV$ of nodes are infected with a rumor, and any other nodes are not infected.
Then, we denote the set of \textit{boundary} nodes which may be infected by the infected nodes $\hat\cV$ by $\cB(\hat \cV)$, i.e., $\cB(\hat \cV) = \{\cup_{v\in \hat \cV}\, \cN(v) \}\backslash \hat \cV$. Let $\cS_{n}$ be the set of ordered $n$ nodes on possible paths of infection, i.e., $\cS_{n} = \{v^n \in \cV^n: v_{i+1}\in \cB(\{v_1,\cdots,v_i\}) \}$, where $v^n=(v_1,v_2,\cdots,v_n)$. Since $\{\tau_{ij}\}$ are independent and these have the memoryless property, an infecting node is uniformly selected from boundary nodes at each step. Hence, we have for any $v^{n - 1}\in \cS_{n - 1}$ and $v_{n}\in \cB(\{v_1,\cdots,v_{n-1}\})$,
\begin{align}
    \Pr\{V_{n} = v_{n}|V^{n-1}=v^{n-1}\}
    &= \frac{1}{|\cB(\{v_1,\cdots,v_{n-1}\})|}\notag\\
    &= \frac{1}{(n-1)\delta-2(n-2)}.
    \label{equ:prob of vn given vn-1}
\end{align}

Let $(v^{(d,0)},v^{(d,1)},\cdots,v^{(d,d)})$ be the (shortest) path from the rumor source $v_1=v^{(d,0)}$ to $v^{(d)}=v^{(d,d)}$. Then, for $d \geq 1$ and $k \geq d + 1$, the $k$th infected node is $v^{(d)}$ if and only if the following event occurs for some $j_1, j_2, \cdots, j_{d}$ such that $2\leq j_1< j_2 < \cdots < j_{d-1} < j_{d} = k$:
\begin{align*}
    &\{V_{j_1} = v^{(d,1)}, V_{j_2} = v^{(d,2)},\cdots,V_{j_d} = v^{(d,d)}\}\\
    &=\{V_2 \neq v^{(d,1)},
    V_3 \neq v^{(d,1)},
    \cdots,
    V_{j_1-1} \neq v^{(d,1)},
    V_{j_1} = v^{(d,1)},\\
    &\quad V_{j_1 +1} \neq v^{(d,2)},
    \cdots,
    V_{j_2 - 1} \neq v^{(d,2)},
    V_{j_2} = v^{(d,2)},
    \cdots,\\
    &\quad V_{j_{d-1}+1} \neq v^{(d,d)},
    \cdots,
    V_{j_{d}-1} \neq v^{(d,d)},
    V_{j_{d}}=v^{(d,d)} \}\\
    &=\cap_{i=1}^{d} \Big\{ \cap_{l=j_{i-1}+1}^{j_i-1}\{V_l \neq v^{(d,i)}\},
    V_{j_i} = v^{(d,i)}\Big\}\\
    &= \cap_{i=1}^{d} \cE_i,
\end{align*}
where $\cE_i = \{ \cap_{l=j_{i-1}+1}^{j_i-1} \{V_l \neq v^{(d,i)}\}, V_{j_i} = v^{(d,i)} \}$ and $j_0=1$. Hence, if $d \geq 2$ and $k \geq d + 1$, we have
\begin{align}
    &\Pr\{V_k=v^{(d)}\}\notag\\
    &= \sum_{2\leq j_1 < j_2<\cdots<j_{d-1} < j_{d} = k}
    \Pr\left\{ \cap_{i=1}^{d}  \cE_i \right\}\notag\\
    &= \sum_{2\leq j_1 < j_2<\cdots<j_{d-1} \leq k-1}
    \Pr\left\{ \cap_{i=1}^{d}  \cE_i \right\}\notag\\
    &\overset{\rm (a)}{=} \sum_{2\leq j_1 < j_2<\cdots<j_{d-1} \leq k-1}
    \prod_{i=1}^{d} \Pr\left\{\cE_i |\cap_{m=1}^{i-1} \cE_m\right\}\notag\\
    &\overset{\rm (b)}{=} \sum_{2\leq j_1 < j_2<\cdots<j_{d-1} \leq k-1}
    \prod_{i=1}^{d} \left(\frac{1}{(j_i-1)\delta - 2 (j_i-2)}\right. \notag\\
    &\quad \left. \times \prod_{l=j_{i-1}+1}^{j_i-1}
        \frac{(l-1)\delta - 2 (l-2)-1}{(l-1)\delta - 2 (l-2)}\right)
    \label{equ:from_app_a}\\
    &= \sum_{2\leq j_1 < j_2<\cdots<j_{d-1} \leq k-1}
    \prod_{i=1}^{d} \bigg(\frac{1}{(j_i-1)\delta - 2 (j_i-2)-1}\notag\\
    &\quad \times \prod_{l=j_{i-1}+1}^{j_i}
    \frac{(l-1)\delta - 2 (l-2)-1}{(l-1)\delta - 2 (l-2)}\bigg)\notag\\
    &= \sum_{2\leq j_1 < j_2<\cdots<j_{d-1} \leq k-1}
    \bigg(\prod_{i=1}^{d} \frac{1}{(j_i-1)\delta - 2 (j_i-2)-1}\bigg)\notag\\
    &\quad \times \bigg( \prod_{i=1}^{d} \prod_{l=j_{i-1}+1}^{j_i}
    \frac{(l-1)\delta - 2 (l-2)-1}{(l-1)\delta - 2 (l-2)}\bigg)\notag\\
    &= \sum_{2\leq j_1 < j_2<\cdots<j_{d-1} \leq k-1}
    \bigg(\prod_{i=1}^{d} \frac{1}{(j_i-1)\delta - 2 (j_i-2)-1}\bigg)\notag\\
    &\quad \times \prod_{l=2}^{j_d}
    \frac{(l-1)\delta - 2 (l-2)-1}{(l-1)\delta - 2 (l-2)}\notag\\
    &= \left(\prod_{l=2}^{k}
        \frac{(l-1)\delta - 2 (l-2)-1}{(l-1)\delta - 2 (l-2)}\right)
    \sum_{2\leq j_1 < j_2<\cdots<j_{d-1} \leq k-1}\notag\\
    &\quad \times  \left(\prod_{i=1}^{d} \frac{1}{(j_i-1)\delta - 2 (j_i-2)-1}\right)\notag\\
    &= \bigg(\prod_{l=2}^{k}
    \frac{(l-1)\delta - 2 (l-2)-1}{(l-1)\delta - 2 (l-2)}\bigg)
    \frac{1}{(k-1)\delta - 2 (k-2)-1}\notag\\
    &\quad \times \sum_{2\leq j_1 < j_2<\cdots<j_{d-1} \leq k-1}
    \bigg(\prod_{i=1}^{d-1} \frac{1}{(j_i-1)\delta - 2 (j_i-2)-1}\bigg)\notag\\
    &= \frac{\prod_{l=2}^{k-1}(l-1)\delta - 2 (l-2)-1}{\prod_{l=2}^{k}(l-1)\delta - 2 (l-2)}
    \sum_{2\leq j_1 < j_2<\cdots<j_{d-1} \leq k-1}\notag\\
    &\quad \times  \bigg(\prod_{i=1}^{d-1} \frac{1}{(j_i-1)\delta - 2 (j_i-2)-1}\bigg)\notag\\
    &= \frac{\prod_{l=2}^{k-1}(l-1)(\delta - 2) + 1}{\prod_{l=2}^{k}(l-1)(\delta - 2)+2}\notag\\
    &\quad \times \sum_{2\leq j_1 < j_2<\cdots<j_{d-1} \leq k-1}
    \bigg(\prod_{i=1}^{d-1} \frac{1}{(j_i-1)(\delta - 2)+1}\bigg)\notag\\
    &=\frac{\prod_{l=2}^{k-1}(\delta-2)\Big(l-1+ \frac{1}{\delta - 2} \Big)}
    {\prod_{l=2}^{k}(\delta - 2)\Big(l-1+\frac{2}{\delta - 2}\Big)}\notag\\
    &\quad \times \sum_{2\leq j_1 < j_2<\cdots<j_{d-1} \leq k-1}
    \bigg(\prod_{i=1}^{d-1} \frac{1}{(\delta - 2)\Big(j_i-1+\frac{1}{\delta - 2}\Big)}\bigg)\notag\\
    &= \frac{(\delta-2)^{k-2}}{(\delta - 2)^{k-1}}
    \frac{\prod_{l=2}^{k-1}\Big(l-1+ \frac{1}{\delta - 2} \Big)}
    {\prod_{l=2}^{k}\Big(l-1+\frac{2}{\delta - 2}\Big)}
    \frac{1}{(\delta - 2)^{d-1}}\notag\\
    &\quad \times\sum_{2\leq j_1 < j_2<\cdots<j_{d-1} \leq k-1}
    \bigg(\prod_{i=1}^{d-1} \frac{1}{j_i-1+\frac{1}{\delta - 2}}\bigg)\notag\\
    &= \frac{1}{(\delta - 2)^{d}}
    \frac{\prod_{l=2}^{k-1}\Big(l-1+ \frac{1}{\delta - 2} \Big)}
    {\prod_{l=2}^{k}\Big(l-1+\frac{2}{\delta - 2}\Big)}\notag\\
    &\quad \times \sum_{2\leq j_1 < j_2<\cdots<j_{d-1} \leq k-1}
    \bigg(\prod_{i=1}^{d-1} \frac{1}{j_i-1+\frac{1}{\delta - 2}}\bigg)\notag\\
    &= \frac{1}{(\delta - 2)^{d}}
    \frac{\prod_{l=1}^{k-2}\Big(l+ \frac{1}{\delta - 2} \Big)}
    {\prod_{l=1}^{k-1}\Big(l+\frac{2}{\delta - 2}\Big)}\notag\\
    &\quad \times \sum_{1\leq j_1 < j_2<\cdots<j_{d-1} \leq k-2}
    \bigg(\prod_{i=1}^{d-1} \frac{1}{j_i+\frac{1}{\delta - 2}}\bigg)\notag\\
    &=\frac{2}{(\delta - 2)^{d}}
    \frac{\frac{1}{\delta-2}\prod_{l=1}^{k-2}\Big(l+ \frac{1}{\delta - 2} \Big)}
    {\frac{2}{\delta-2}\prod_{l=1}^{k-1}\Big(l+\frac{2}{\delta - 2}\Big)}\notag\\
    &\quad \times \sum_{1\leq j_1 < j_2<\cdots<j_{d-1} \leq k-2}
    \bigg(\prod_{i=1}^{d-1} \frac{1}{j_i+\frac{1}{\delta - 2}}\bigg)\notag\\
    &= \frac{2}{(\delta - 2)^{d}}
    \frac{\prod_{l=0}^{k-2}\Big(l+ \frac{1}{\delta - 2} \Big)}
    {\prod_{l=0}^{k-1}\Big(l+\frac{2}{\delta - 2}\Big)}\notag\\
    &\quad \times \sum_{1\leq j_1 < j_2<\cdots<j_{d-1} \leq k-2}
    \bigg(\prod_{i=1}^{d-1} \frac{1}{j_i+\frac{1}{\delta - 2}}\bigg)\notag\\
    &= \frac{2}{(\delta - 2)^{d}}
    \frac{\Big(\frac{1}{\delta - 2} \Big)^{\overline{k-1}}}
    {\Big(\frac{2}{\delta - 2}\Big)^{\overline{k}}}
    \zeta_{k-2}^{d-1}\Big(\frac{1}{\delta-2}\Big)\notag\\
    &= p_{1}(\delta, d, k),
    \label{equ:prob vk}
\end{align}
where (a) comes from the chain rule of the probability, and (b) comes from Appendix \ref{app:a}.

The remaining case is that $d = 1$ and $k \geq d + 1\, (= 2)$. In this case, we have
\begin{align}
    &\Pr\{V_k=v^{(d)}\}\notag\\
    &= \Pr\left\{ \cE_1 \right\}\notag\\
    &\overset{\rm (a)}{=}\left(\frac{1}{(j_1-1)\delta - 2 (j_1-2)} \notag \right.\\
    &\quad \left. \times \prod_{l=j_{0}+1}^{j_1-1} \frac{(l-1)\delta - 2 (l-2)-1}{(l-1)\delta - 2 (l-2)}\right)\notag\\
    &= \left(\frac{1}{(j_1-1)\delta - 2 (j_1-2) - 1} \notag \right.\\
    &\quad \left. \times \prod_{l=j_{0}+1}^{j_1} \frac{(l-1)\delta - 2 (l-2)-1}{(l-1)\delta - 2 (l-2)}\right)\notag\\
    &= \left(\frac{1}{(k-1)\delta - 2 (k-2) - 1}  \prod_{l= 2}^{k} \frac{(l-1)\delta - 2 (l-2)-1}{(l-1)\delta - 2 (l-2)}\right)\notag\\
    &= \left(\frac{1}{(k-1)(\delta - 2) + 1}  \prod_{l= 2}^{k} \frac{(l-1)(\delta - 2) + 1}{(l-1)(\delta - 2) + 2}\right)\notag\\
    &= \left(\frac{2}{(k-1)(\delta - 2) + 1}  \prod_{l= 1}^{k} \frac{(l-1)(\delta - 2) + 1}{(l-1)(\delta - 2) + 2}\right)\notag\\
    &=  \left(\frac{2}{\delta - 2} \frac{1}{k - 1+ \frac{1}{\delta - 2}}  \prod_{l= 1}^{k} \frac{l - 1 + \frac{1}{\delta - 2}}{l - 1 + \frac{2}{\delta - 2}}\right)\notag\\
    &=  \left(\frac{2}{\delta - 2} \frac{1}{k - 1+ \frac{1}{\delta - 2}}  \prod_{l= 0}^{k - 1} \frac{l + \frac{1}{\delta - 2}}{l + \frac{2}{\delta - 2}}\right)\notag\\
    &= \frac{2}{\delta - 2} \frac{ \prod_{l= 0}^{k-2} \left(l + \frac{1}{\delta - 2} \right)}{\prod_{l= 0}^{k - 1} \left(l + \frac{2}{\delta - 2} \right)}\notag\\
    &= \frac{2}{\delta - 2} \frac{\left(\frac{1}{\delta - 2} \right)^{\overline{k - 1}}}{\left(\frac{2}{\delta - 2} \right)^{\overline{k}}},
    \label{equ:prob_vk_for1}
\end{align}
where (a) comes from Appendix \ref{app:a}. Thus, by recalling that $\zeta_{k-2}^{d-1}\Big(\frac{1}{\delta-2}\Big) = 1$ if $d = 1$ and $k \geq 2$, \eqref{equ:prob_vk_for1} implies that \eqref{equ:prob vk} also holds in this case.

Consequently, \eqref{equ:prob vk} holds for any $d \geq 1$ and $k \geq d + 1$.

\subsection{Closed-Form Expression of $\Pr\{X^\delta(v^{(d)}) = x^\delta| V_k=v^{(d)}\}$}
Suppose that the $k$th infected node is $v_{k}$. Since we consider a regular tree, $v_{k}$ has $\delta$ neighboring nodes $\{v_{k,1},\cdots,v_{k,\delta}\}$. Let $Y_j(v_k)$ be the number of infected nodes of the subtree rooted at $v_{k,j}$ with the parent node $v_{k}$ after $v_k$ is infected. Let the subtree rooted at $v_{k, \delta}$ contain the rumor source. Thus, at the time that $v_k$ is infected, it holds that $X_{\delta}(v_k) = k - 1$. From then on, an infecting node is uniformly selected from boundary nodes at each step. We note that $X_j(v_k)=Y_j(v_k)$ for all $j\in\{1,2,\cdots,\delta-1\}$, and $X_\delta(v_k)=Y_\delta(v_k)+k-1$.
Then, numbers $\{Y_j(v_k)\}$ are drawn according to the P\'olya's urn model with $\delta$ colors balls
(cf. \cite{shah2011rumors} and \cite{dong2013rootingArXiv4}):
Initially, $b_j$ balls of color $C_j$ $(j=1,2,\cdots,\delta)$ are in the urn, where $b_j = 1$ if $j\neq \delta$ and $b_j = (k-1)(\delta-2) + 1$ if $j = \delta$. At each step, a single ball is uniformly drawn form the urn. Then, the drawn ball is returned with additional $m = \delta - 2$ balls of the same color. Repeat this drawing process.

$Y_j(v_k)$ corresponds to the number of times that the balls of color $C_j$ are drawn.
According to \cite[Chap.\ 4]{johnson1977urn}, when the total number of drawing balls is $n-k$, the joint distribution of $Y^\delta(v_k)=(Y_{1}(v_k),\cdots,Y_{\delta}(v_k))$ is given by
\begin{align}
    \Pr&\{Y^\delta(v_k)=y^\delta\}\notag\\
    =& \frac{(n-k)!}{y_1!\cdots y_\delta !}\frac{\prod_{j=1}^\delta b_j(b_j+m)\cdots(b_j+(y_j-1)m)}{b(b+m)\cdots(b+(n-k-1)m)},
    \label{equ:joint prob of Polya's Urn model}
\end{align}
where $b = \sum_{j=1}^\delta b_j$ and $\sum_{j=1}^\delta y_j =n-k$.
We note that the above probability only depends on $n$, $k$ and $\delta$.

Now, by definition, we have 
\begin{align}
    &\Pr\{ X^\delta(v^{(d)}) = x^\delta| V_k=v^{(d)} \}\notag\\
    &= \Pr\{Y^{\delta}(v^{(d)}) = (x_1, x_2, \cdots, x_{\delta-1}, x_\delta - k + 1)\}.
    \label{equ:relatoin to X and Y}
\end{align}

\subsection{Proof of Theorem \ref{thm:asymptotic delta=3}}
When $\delta=3$, according to (\ref{equ:prob vk}), (\ref{equ:joint prob of Polya's Urn model}) and (\ref{equ:relatoin to X and Y}), we have
\begin{align}
    \Pr\{V_k=v^{(d)}\}
    &= \frac{2}{(k+1)k}\zeta_{k-2}^{d-1}(1),
    \label{equ:prob vk for deg 3}\\
    \Pr\{X^3(v^{(d)}) = x^3| V_k=v^{(d)} \}
    & = \frac{\binom{x_3}{k-1}}{\binom{n+1}{k+1}},
    \label{equ:prob x cond vk for deg 3}
\end{align}
for any $d \geq 1$ and $k \geq d + 1$.

When $n$ is odd, we have $\Pr\{v^{(d)}\in\cV_n, \overline X(v) = n/2\} = 0$. Thus, we only consider the first term of (\ref{equ:distance prob with lemma}). According to (\ref{equ:prob vk}), (\ref{equ:joint prob of Polya's Urn model}) and (\ref{equ:relatoin to X and Y}), (\ref{equ:correct prob < n/2}) can be represented as
\begin{align*}
    &\sum_{k=d+1}^{\lceil n/2 \rceil}\frac{2}{(k+1)k}\zeta_{k-2}^{d-1}(1)
    \sum_{x^3:\sum_{i=1}^3 x_i=n-1, \atop \max_{1\leq j \leq 3}\{x_j\} < n/2}\frac{\binom{x_3}{k-1}}{\binom{n+1}{k+1}}\\
    &= \sum_{k=d+1}^{(n+1)/2}
    \frac{2}{k+1} \zeta_{k-2}^{d-1}(1)
    \frac{\binom{(n+3)/2}{k+1}}
    {\binom{n+1}{k+1}},
\end{align*}
where the equality follows since
\begin{align*}
    \sum_{x^3:\sum_{i=1}^3 x_i=n-1, \atop \max_{1\leq j \leq 3}\{x_j\} < n/2} \binom{x_3}{k-1}
    &= \sum_{x_3=k-1}^{(n-1)/2}\sum_{x^2:\sum_{i=1}^3 x_i=n-1,\atop \max_{1\leq j \leq 3}\{x_j\} < n/2} \binom{x_3}{k-1}\\
    &= \sum_{x_3=k-1}^{(n-1)/2}\sum_{x_2=(n-1)/2-x_3}^{(n-1)/2} \binom{x_3}{k-1}\\
    &= \sum_{x_3=k-1}^{(n-1)/2}(x_3+1) \binom{x_3}{k-1}\\
    &= k \sum_{x_3=k-1}^{(n-1)/2}\binom{x_3+1}{k}\\
    &= k \binom{(n+3)/2}{k+1}.
\end{align*}
Thus, we have
\begin{align*}
    \bfD_{n}(d)
    = 3\cdot 2^{d-1}
    \sum_{k=d+1}^{(n+1)/2}
    \frac{2}{k+1} \zeta_{k-2}^{d-1}(1)
    \frac{\binom{(n+3)/2}{k+1}}
    {\binom{n+1}{k+1}}.
\end{align*}

In a similar way, we have $\bfD_{n}(d)$ for even $n$ as follows:
\begin{align*}
    \bfD_{n}(d)
    &=
    3\cdot 2^{d-1}
    \sum_{k=d+1}^{n/2+1}
    \frac{2}{k+1} \zeta_{k-2}^{d-1}(1)\\
    &\quad \times \frac{\binom{n/2+1}{k+1}
      +  \frac{n}{2(n+2)}
      \binom{n/2+1}{k}}{\binom{n+1}{k+1}}.
\end{align*}

This is because
\begin{align*}
    & \sum_{x^3:\sum_{i=1}^3 x_i=n-1, \atop \max_{1\leq j \leq 3}\{x_j\} <n/2} \binom{x_3}{k-1}
    +\frac{1}{2} \sum_{x^3:\sum_{i=1}^3 x_i=n-1, \atop \max_{1\leq j \leq 3}\{x_j\} =n/2} \binom{x_3}{k-1}\\
    &\eqo{(a)} \sum_{x_3=k-1}^{n/2-1}x_3\binom{x_3}{k-1}
    +\sum_{x_3=k-1}^{n/2-1}\binom{x_3}{k-1} + \frac{n}{4} \binom{n/2}{k-1}\\
    &=\left(\sum_{x_3=k-1}^{n/2-1}(x_3+1)\binom{x_3}{k-1}\right)
    + \frac{n}{4}\frac{k}{n/2+1} \binom{n/2+1}{k}\\
    &=\left(\sum_{x_3=k-1}^{n/2-1}(x_3+1)\binom{x_3}{k-1}\right)
    + k \frac{n}{4}\frac{2}{n+2} \binom{n/2+1}{k}\\
    &=k\left(\sum_{x_3=k-1}^{n/2-1}\binom{x_3+1}{k}\right)
    + k \frac{n}{2(n+2)} \binom{n/2+1}{k}\\
    &=k\binom{n/2+1}{k+1}
    + k \frac{n}{2(n+2)} \binom{n/2+1}{k},
\end{align*}
where (a) follows since
\begin{align*}
    &\sum_{x^3:\sum_{i=1}^3 x_i=n-1, \atop \max_{1\leq j \leq 3}\{x_j\} <n/2} \binom{x_3}{k-1}\\
    &= \sum_{x_3=k-1}^{n/2-1}\sum_{x^2:\sum_{i=1}^3 x_i=n-1,\atop \max_{1\leq j \leq 3}\{x_j\} < n/2} \binom{x_3}{k-1}\\
    &= \sum_{x_3=k-1}^{n/2-1}\sum_{x_2=n/2-x_3}^{n/2-1} \binom{x_3}{k-1}\\
    &= \sum_{x_3=k-1}^{n/2-1}x_3\binom{x_3}{k-1},
\end{align*}
and
\begin{align*}
    &\sum_{x^3:\sum_{i=1}^3 x_i=n-1, \atop \max_{1\leq j \leq 3}\{x_j\} =n/2} \binom{x_3}{k-1}\\
    &= \sum_{x_3=k-1}^{n/2}\sum_{x^2:\sum_{i=1}^3 x_i=n-1,\atop \max_{1\leq j \leq 3}\{x_j\} = n/2} \binom{x_3}{k-1}\\
    &= \sum_{x_3=k-1}^{n/2 - 1} \sum_{x^2:\sum_{i=1}^3 x_i=n-1,\atop \max_{1\leq j \leq 3}\{x_j\} = n/2} \binom{x_3}{k-1}\\
    &\quad + \sum_{x^2:\sum_{i=1}^3 x_i=n-1,\atop \max_{1\leq j \leq 3}\{x_j\} = n/2, x_3 = n/2} \binom{n/2}{k-1}\\
    &= \sum_{x_3=k-1}^{n/2 - 1} \sum_{x^2:\sum_{i=1}^3 x_i=n-1,\atop \max_{1\leq j \leq 2}\{x_j\} = n/2} \binom{x_3}{k-1}\\
    &\quad + \sum_{x^2: x_1 + x_2 = n/2 - 1} \binom{n/2}{k-1}\\
    &= \sum_{x_3=k-1}^{n/2-1}2 \binom{x_3}{k-1} + \frac{n}{2} \binom{n/2}{k-1}.
\end{align*}

Since $\zeta_k^d(1)  = \zeta_{k+1}^{d}(0) - \zeta_{k}^{d-1}(1)$ for any $d \geq 1$ and $k \geq d$ (see Appendix \ref{app:zeta_function}), we have for any $d \geq 2$ and $k \geq d + 1$,
\begin{align*}
    \zeta_{k-2}^{d-1}(1)
    =& \zeta_{k-1}^{d-1}(0) - \zeta_{k-2}^{d-2}(1)\\
    =& \zeta_{k-1}^{d-1}(0) - \zeta_{k-1}^{d-2}(0) + \zeta_{k-2}^{d-3}(1)\\
    =& \zeta_{k-1}^{d-1}(0) - \zeta_{k-1}^{d-2}(0) + \zeta_{k-1}^{d-3}(0) - \zeta_{k-2}^{d-4}(1)\\
    \vdots\\
    =& \sum_{l=1}^{d} (-1)^{d-l} \zeta_{k-1}^{l-1}(0),
\end{align*}
where $\zeta_{k-1}^{0}(0) = 1$. Note that this holds even if $d = 1$ and $k \geq d+ 1$. Since it holds \cite{adamchik1997stirling,kuba2010note} that
\begin{align*}
    \zeta_{k-1}^{l-1}(0) = \frac{1}{(k-1)!}\left[{k \atop l}\right]
\end{align*}
for any $l \geq 1$ and $k \geq l$, we have for any $d \geq 1$ and $k \geq d + 1$,
\begin{align*}
    \zeta_{k-2}^{d-1}(1)
    &= \sum_{l=1}^{d} (-1)^{d-l} \zeta_{k-1}^{l-1}(0)\\
    &= \sum_{l=1}^{d} (-1)^{d-l} \frac{1}{(k-1)!}\left[{k \atop l}\right]\\
    &= (-1)^{d-k}\frac{1}{(k-1)!}\sum_{l=1}^{d} (-1)^{k-l} \left[{k \atop l}\right]\\
    &= \frac{(-1)^{d+k}}{(k-1)!}\sum_{l=1}^{d} s(k,l),
\end{align*}
where  $[{k \atop l}]$ is the {\it unsigned} Stirling numbers of the first kind \cite{graham1994concrete} and $s(k,l)$ is the {\it signed} Stirling numbers of the first kind \cite{graham1994concrete} defined as $s(k,l) \teq (-1)^{k-l}\left[{k \atop l}\right]$. Thus, we have for odd $n \geq 3$,
\begin{align}
    \bfD_{n}(d) = 3\cdot 2^{d-1}
    \sum_{k=d+1}^{(n+1)/2}
    \frac{2}{k+1} 
    \frac{\binom{(n+3)/2}{k+1}}
    {\binom{n+1}{k+1}}
    \frac{(-1)^{d+k}}{(k-1)!}\sum_{l=1}^{d} s(k,l),
    \label{equ:D_for_odd_with_str_num}
\end{align}
and for even $n \geq 2$,
\begin{align}
    \bfD_{n}(d) &= 3\cdot 2^{d-1}
    \sum_{k=d+1}^{n/2+1}
    \frac{2}{k+1} 
    \frac{\binom{n/2+1}{k+1}
      +  \frac{n}{2(n+2)}
      \binom{n/2+1}{k}}{\binom{n+1}{k+1}}\notag\\
    &\quad \times   \frac{(-1)^{d+k}}{(k-1)!}\sum_{l=1}^{d} s(k,l).
    \label{equ:D_for_even_with_str_num}
\end{align}

Now, the well-known Lebesgue's dominated convergence theorem and the fact that
\begin{align*}
    \lim_{m\ra\infty}\frac{\binom{((2 m + 1) + 3)/2}{k+1} }{\binom{(2 m + 1)+1}{k+1}}
    &= \lim_{m\ra\infty}\frac{\binom{m + 2}{k+1} }{\binom{2 m + 2}{k+1}}\\
    &= \frac{1}{2^{k+1}}
\end{align*}
and
\begin{align*}
    &\lim_{m\ra\infty} \frac{\binom{(2 m)/2+1}{k+1} + \frac{(2 m)}{2((2 m)+2)} \binom{(2 m)/2+1}{k}}{\binom{(2 m)+1}{k+1}}\\
    &\lim_{m\ra\infty} \frac{\binom{m +1}{k+1} + \frac{m}{2(m+1)} \binom{m+1}{k}}{\binom{2 m+1}{k+1}}\\
    &= \frac{1}{2^{k+1}},
\end{align*}
implies (see a precise derivation in Appendix \ref{app:limit_of_Dn})
\begin{align}
    \lim_{n\ra\infty}
    \bfD_{n}(d)
    = 3\cdot 2^{d-1}
    \sum_{k=d+1}^{\infty}
    \frac{1}{2^{k}}
    \frac{1}{k+1}
    \frac{ (-1)^{d+k}}{(k-1)!}\sum_{l=1}^{d} s(k,l).
    \label{equ:limit_of_Dn}
\end{align}
Thus, we can evaluate the probability as follows:
\begin{align*}
    &\lim_{n\ra\infty} \bfD_{n}(d)\\
    &= 3\cdot 2^{d-1}
    \sum_{k=d+1}^{\infty}
    \frac{ (-1)^{d+k}}{2^{k}}
    \frac{1}{k+1}
    \frac{1}{(k-1)!}\sum_{l=1}^{d} s(k,l)\\
    &= 3\cdot 2^{d-1}
    (-1)^{d}
    \sum_{l=1}^{d}
    \sum_{k=d+1}^{\infty}
    \frac{k}{k+1}
    \frac{(-\frac{1}{2})^k}{k!}
    s(k,l)\\
    &= 3\cdot 2^{d-1}
    (-1)^{d}
    \sum_{l=1}^{d}
    \left(
        - \sum_{k=l}^{d}
        \frac{k}{k+1}
        \frac{(-\frac{1}{2})^k}{k!}
        s(k,l) \right.\\
    &\left.\quad        
        + \sum_{k=l}^{\infty}
        \frac{k}{k+1}
        \frac{(-\frac{1}{2})^k}{k!}
        s(k,l)
    \right)\\
    &\overset{\rm (a)}{=} 3\cdot 2^{d-1}
    (-1)^{d}
    \sum_{l=1}^{d}
    \left(
        - \sum_{k=l}^{d}
        \frac{k}{k+1}
        \frac{(-\frac{1}{2})^k}{k!}
        s(k,l)        
    \right.\\
    &\left.\quad + (-1)^l\left( \frac{(\ln 2)^l}{l!} -2 + \sum_{m=0}^l \frac{(\ln 2)^m}{m!} \right) \right)\\
    &= 3\cdot 2^{d-1}
    (-1)^{d}
    \left(
        -  \sum_{l=1}^{d} \sum_{k=l}^{d}
        \frac{k}{k+1}
        \frac{(-\frac{1}{2})^k}{k!}
        s(k,l)
    \right.\\
    &\quad \left.
        + \sum_{l=1}^{d}
        (-1)^l\left( \frac{(\ln 2)^l}{l!} -2 + \sum_{m=0}^l \frac{(\ln 2)^m}{m!} \right)
    \right)\\
    &= 3\cdot 2^{d-1}
    (-1)^{d}
    \left(
        - \sum_{k=1}^{d} \sum_{l=1}^{k} 
        \frac{k}{k+1}
        \frac{(-\frac{1}{2})^k}{k!}
        s(k,l)
    \right.\\
    &\quad \left.
        + \sum_{l=1}^{d}
        (-1)^{l}\left( \frac{(\ln 2)^l}{l!} -2 + \sum_{m=0}^l \frac{(\ln 2)^m}{m!} \right)
    \right)\\
    &= 3\cdot 2^{d-1}
    (-1)^{d}
    \left(
        - \sum_{k=1}^{d} 
        \frac{k}{k+1}
        \frac{(-\frac{1}{2})^k}{k!}
        \sum_{l=1}^{k} 
        s(k,l)
    \right.\\
    &\quad \left.
        + \sum_{l=1}^{d}
        (-1)^l \left( \frac{(\ln 2)^l}{l!} -2 + \sum_{m=0}^l \frac{(\ln 2)^m}{m!} \right)
    \right)\\
    &\overset{\rm (b)}{=} 3\cdot 2^{d-1}
    (-1)^{d}
    \left(
        - \frac{1}{2}
        \frac{(-\frac{1}{2})}{1}
    \right.\\
    &\quad \left.
        + \sum_{l=1}^{d}
        (-1)^l\left( \frac{(\ln 2)^l}{l!} -2 + \sum_{m=0}^l \frac{(\ln 2)^m}{m!} \right)
    \right)\\
    &= 3\cdot 2^{d-1}
    (-1)^{d}
    \left(
        \frac{1}{4}
    \right.
    \\
    &\quad \left.
        + \sum_{l=1}^{d}
        (-1)^l \left( \frac{(\ln 2)^l}{l!} -2 + \sum_{m=0}^l \frac{(\ln 2)^m}{m!} \right)
    \right)\\
    &= f(d),
\end{align*}
where (a) comes from Appendix \ref{app:bivariate_GF_SN1}, and (b) follows since $\sum_{l=1}^{k} s(k,l) = 1$ if $k = 1$ and $\sum_{l=1}^{k} s(k,l) = 0$ if $k \neq 1$. This completes the proof of Theorem \ref{thm:asymptotic delta=3}.
\subsection{Proof of Theorem \ref{thm:lower upper any delta}}
In this section, we denote $I_{1/2}(k-1+\frac{\delta-1}{\delta-2},\frac{1}{\delta-2})$ by $I^{(1)}(\delta, k)$ and $I_{1/2}(k-1+\frac{1}{\delta-2}, \frac{\delta-1}{\delta-2})$ by $I^{(2)}(\delta, k)$.

Let $\cE_{j}(v^{(d)}) \teq \{X_j(v^{(d)}) < n/2\}$. Due to (\ref{equ:distance prob with lemma}), we have
\begin{align}
    &\Pr\{v^{(d)}\in\cV_n, \hat V_{n} = v^{(d)} \}\notag\\
    &\geq \Pr\big\{v^{(d)}\in\cV_n, \overline X(v^{(d)})< n/2\big\}\notag\\
    &\geq \Pr\{v^{(d)}\in\cV_n, \cap_{j=1}^\delta \cE_{j}(v^{(d)}) \}\notag\\
    &= \sum_{k=d+1}^{\lceil n/2 \rceil} \Pr\big\{V_k=v^{(d)}, \cap_{j=1}^\delta \cE_{j}(v^{(d)}) \big\}\notag\\
    &= \sum_{k=d+1}^{\lceil n/2 \rceil} \Pr\big\{V_k=v^{(d)}\big\}
    \Pr\big\{\cap_{j=1}^\delta \cE_{j}(v^{(d)}) | V_k=v^{(d)}\big\}\notag\\
    &\geqo{(a)} \sum_{k=d+1}^{\lceil n/2 \rceil} \Pr\big\{V_k=v^{(d)}\big\}
    \Big(\Pr\big\{\cE_{\delta}(v^{(d)}) | V_k=v^{(d)}\big\}\notag\\
    &\quad -\Pr\big\{\cup_{j=1}^{\delta-1} [\cE_{j}(v^{(d)})]^c | V_k=v^{(d)}\big\} \Big)\notag\\
    &\geq \sum_{k=d+1}^{\lceil n/2 \rceil} \Pr\big\{V_k=v^{(d)}\big\}
    \Big(\Pr\big\{\cE_{\delta}(v^{(d)}) | V_k=v^{(d)}\big\}\notag\\
    &\quad -\sum_{j=1}^{\delta-1} \Pr\big\{[\cE_{j}(v^{(d)})]^c | V_k=v^{(d)}\big\} \Big)\notag\\
    &\eqo{(b)} \sum_{k=d+1}^{\lceil n/2 \rceil} \Pr\{V_k=v^{(d)}\} \left( \Pr\{ \cE_{\delta}(v^{(d)}) | V_k=v^{(d)}\} \right. \notag\\
    &\quad \left. - (\delta-1) (1-\Pr\{\cE_{1}(v^{(d)}) | V_k=v^{(d)}\}) \right),
    \label{equ:upper bound of dist prob using E}
\end{align}
where (a) comes from the fact that
\begin{align*}
    \Pr\{\cap_{i=1}^m \cA_i\}
    \geq \Pr\{\cA_m\} - \Pr\{\cup_{i=1}^{m-1} \cA_i^c\},
\end{align*}
and (b) comes from the symmetric property of $\cE_{i}(v^{(d)})$ for all $1\leq i \leq \delta-1$.
Similarly, by letting $\cF_{j}(v^{(d)}) \teq \{X_j(v^{(d)}) \leq n/2\}$, we have
\begin{align}
    &\Pr\{v^{(d)}\in\cV_n, \hat V_{n} = v^{(d)} \}\notag\\
    &\leq \Pr\big\{v^{(d)}\in\cV_n, \overline X(v^{(d)})\leq n/2\big\}\notag\\
    &\leq \Pr\{ \cup_{k=d+1}^{n} V_k=v^{(d)}, \cap_{i=1}^\delta \cF_{i}(v^{(d)})\}\notag\\
    &= \sum_{k=d+1}^{\lfloor n/2 \rfloor + 1} \Pr\{V_k=v^{(d)}, \cap_{i=1}^\delta \cF_{i}(v^{(d)})\}\notag\\
    &= \sum_{k=d+1}^{\lfloor n/2 \rfloor + 1} \Pr\{V_k=v^{(d)}\} \Pr\{ \cap_{i=1}^\delta \cF_{i}(v^{(d)})| V_k=v^{(d)}\}\notag\\
    &= \sum_{k=d+1}^{\lfloor n/2 \rfloor + 1} \Pr\{V_k=v^{(d)}\}\notag\\
    &\quad \times \bigg(1-\Pr\{ \cup_{i=1}^\delta [\cF_{i}(v^{(d)})]^c| V_k=v^{(d)}\}\bigg)\notag\\
    &\eqo{(a)} \sum_{k=d+1}^{\lfloor n/2 \rfloor + 1} \Pr\{V_k=v^{(d)}\}\notag\\
    &\quad \times \bigg(1-\sum_{i=1}^\delta \Pr\{ [\cF_{i}(v^{(d)})]^c| V_k=v^{(d)}\}\bigg)\notag\\
    &=  \sum_{k=d+1}^{\lfloor n/2 \rfloor + 1} \Pr\{V_k=v^{(d)}\} \left( \Pr\{ \cF_{\delta}(v^{(d)}) | V_k=v^{(d)}\} \right. \notag\\
    &\quad \left. - (\delta-1) (1-\Pr\{\cF_{1}(v^{(d)}) | V_k=v^{(d)}\}) \right),
    \label{equ:upper bound of dist prob using F}
\end{align}
where (a) comes from the fact that events $[\cF_{1}(v^{(d)})]^c$, $[\cF_{2}(v^{(d)})]^c$, $\cdots$, $[\cF_{\delta}(v^{(d)})]^c$ are disjoint.

By using the same way as in \cite[Chapter III.B]{dong2013rootingArXiv4} (see also \cite[Section 4.1.5]{shah2016finding}), we have (see a precise derivation in Appendix \ref{app:polya})
\begin{align}
    \lim_{n\ra\infty}\Pr\{\cE_{1}(v^{(d)}) | V_k=v^{(d)}\}
    &= \lim_{n\ra\infty}\Pr\{\cF_{1}(v^{(d)}) | V_k=v^{(d)}\}\notag\\
    &= 1 - I^{(1)}(\delta, k),
    \label{equ:F1=E1=I} \\
    \lim_{n\ra\infty}\Pr\{\cE_{\delta}(v^{(d)}) | V_k=v^{(d)}\}
    &= \lim_{n\ra\infty}\Pr\{\cF_{\delta}(v^{(d)}) | V_k=v^{(d)}\}\notag\\
    &= I^{(2)}(\delta, k).
    \label{equ:Fd=Ed=I}
\end{align}

According to these equalities, (\ref{equ:upper bound of dist prob using E}), (\ref{equ:upper bound of dist prob using F}), and the dominated convergence theorem, we have (see a precise derivation in Appendix \ref{app:E=F})
\begin{align}
    \lim_{n\ra\infty}\bfD_{n}(d)
    &= \delta(\delta-1)^{d-1} \sum_{k=d+1}^{\infty} p_{1}(\delta, d, k) \notag\\
    &\quad \times ( I^{(2)}(\delta, k) - (\delta-1)I^{(1)}(\delta, k) )
    \label{equ:dist prob with lim for any degree}\\
    &\geq g(\delta,d,m),\quad \forall m \geq d + 1,\notag
\end{align}
where $g(\delta,d,m)$ is a partial sum of (\ref{equ:dist prob with lim for any degree}),
and the inequality comes from the fact that (according to \eqref{equ:upper bound of dist prob using F}, \eqref{equ:F1=E1=I}, and \eqref{equ:Fd=Ed=I})
\begin{align*}
    0&\leq \lim_{n \to \infty}\Pr\{V_k=v^{(d)}, \cap_{i=1}^\delta \cF_{i}(v^{(d)})\}\\
    &= \lim_{n \to \infty}  \Pr\{V_k=v^{(d)}\} \left( \Pr\{ \cF_{\delta}(v^{(d)}) | V_k=v^{(d)}\} \right. \notag\\
    &\quad \left. - (\delta-1) (1-\Pr\{\cF_{1}(v^{(d)}) | V_k=v^{(d)}\}) \right)\\
    &= p_{1}(\delta,d,k) ( I^{(2)}(\delta, k) - (\delta-1)I^{(1)}(\delta, k) ), \quad \forall k \geq d + 1.
\end{align*}

On the other hand, we have
\begin{align*}
    &g(\delta,d,l) - g(\delta,d,l-1)\\
    &= \delta(\delta-1)^{d-1} p_{1}(\delta, d, l)
    ( I^{(2)}(\delta, l) - (\delta-1)I^{(1)}(\delta, l) )\\
    &\overset{\rm (a)}{\leq}  I^{(2)}(\delta, l) - (\delta-1)I^{(1)}(\delta, l)\\
    &\leq  I^{(2)}(\delta, l)\\
    &=I_{1/2}\left(l-1+\frac{1}{\delta-2}, \frac{\delta-1}{\delta-2}\right)\\
    &\overset{\rm (b)}{\leq} 4 e^2 l(l+1) 2^{- l + 1},
\end{align*}
where (a) comes from the fact that
\begin{align*}
    1 &\geq \Pr\left\{\cup_{v^{(d)} \in \cV^{(d)}} \{V_l=v^{(d)}\}\right\}\\
    &= \sum_{v^{(d)} \in \cV^{(d)}} \Pr\{V_l=v^{(d)}\}\\
    &=  \delta(\delta-1)^{d-1} p_{1}(\delta,d,l),
\end{align*}
and (b) comes from the same (but a bit improved) inequality in \cite[Sect.\ 4.5]{shah2016finding}. Thus, for any $M\geq m+1$, we have
\begin{align*}
    g(\delta,d,M) - g(\delta,d,m)
    &= \sum_{l=m+1}^{M} \left( g(\delta,d,l)  - g(\delta,d,l-1)\right) \\
    &\leq \sum_{l=m+1}^{M} 4 e^2 l(l+1) 2^{- l + 1}.
\end{align*}
Since $\lim_{M\ra\infty}g(\delta,d,M) = \lim_{n\ra\infty} \bfD_n(d)$,
we have
\begin{align*}
    0 &\leq \lim_{n\ra\infty} \bfD_n(d)  - g(\delta,d,m)\\
    & = \lim_{M\ra\infty}g(\delta,d,M)  - g(\delta,d,m)\\
    & = \lim_{M\ra\infty} \left(g(\delta,d,M)  - g(\delta,d,m) \right)\\
    &\leq \sum_{l=m+1}^{\infty} 4 e^2 l(l+1) 2^{- l + 1}\\
    &= e^2 (8 + 5 m + m^2) 2^{- m + 3}, \quad \forall m \geq d + 1.
\end{align*}
This completes the proof of Theorem \ref{thm:lower upper any delta}.

\appendices
\section{} \label{app:a}
We have
\begin{align}
    &\Pr\{\cE_i |\cap_{m=1}^{i-1} \cE_m\}\notag\\
    &=\Pr\Big\{ \cap_{l=j_{i-1}+1}^{j_i-1}\{V_l \neq v^{(d,i)}\},
    V_{j_i} = v^{(d,i)} \big|\cap_{m=1}^{i-1} \cE_m\Big\}\notag\\
    &=\Pr\bigg\{V_{j_i} = v^{(d,i)} \bigg| \bigcap_{l=j_{i-1}+1}^{j_i-1}\{V_l \neq v^{(d,i)}\}, \bigcap_{m=1}^{i-1} \cE_m\bigg\}\notag\\
    &\quad \times \Pr\bigg\{ \bigcap_{l=j_{i-1}+1}^{j_i-1}\{V_l \neq v^{(d,i)}\} \bigg| \bigcap_{m=1}^{i-1} \cE_m\bigg\}\notag\\
    &= \Pr\bigg\{V_{j_i} = v^{(d,i)} \bigg| \bigcap_{l=j_{i-1}+1}^{j_i-1}\{V_l \neq v^{(d,i)}\}, \bigcap_{m=1}^{i-1} \cE_m\bigg\}\notag\\
    &\quad \times \Pr\bigg\{\{V_{j_{i-1}+1} \neq v^{(d,i)}\},\cdots, \{V_{j_i-1} \neq v^{(d,i)}\}\bigg| \bigcap_{m=1}^{i-1} \cE_m\bigg\}\notag\\
    &= \Pr \{V_{j_i} = v^{(d,i)} | \cap_{l=j_{i-1}+1}^{j_i-1} \{V_l \neq v^{(d,i)}\}, \cap_{m=1}^{i-1} \cE_m \}\notag\\
    &\quad \times \prod_{l=j_{i-1}+1}^{j_i-1} \Pr \{ V_l \neq v^{(d,i)} |\cap_{m=j_{i-1}+1}^{l-1} \{V_m \neq v^{(d,i)}\}, \notag \\
    &\quad \cap_{m=1}^{i-1} \cE_m \},
    \label{equ:prob of event Ei}
\end{align}
where we use the converntion that if $j_{i} = j_{i - 1} + 1$,
\begin{align*}
    &\prod_{l=j_{i-1}+1}^{j_i-1} \Pr \{ V_l \neq v^{(d,i)} |\cap_{m=j_{i-1}+1}^{l-1} \{V_m \neq v^{(d,i)}\}, \cap_{m=1}^{i-1} \cE_m \}\\
    &= 1.
\end{align*}
On the other hand, we have
\begin{align}
    &\Pr\{ V_l \neq v^{(d,i)} |\cap_{m=j_{i-1}+1}^{l-1} \{V_m \neq v^{(d,i)}\},  \cap_{m=1}^{i-1} \cE_m \}\notag\\
    &= \Pr\{ V_l \neq v^{(d,i)} |
    \cup_{v^{l-1}\in\cP_{l,i}}
    \{V^{l-1}= v^{l-1}\}\}\notag\\
    &= \frac{\Pr\{ V_l \neq v^{(d,i)},
      \bigcup_{v^{l-1}\in\cP_{l,i}} \{
      V^{l-1}= v^{l-1}\} \}}{\Pr\{ \bigcup_{v^{l-1}\in\cP_{l,i}} \{
      V^{l-1}= v^{l-1}\} \}}\notag\\
    &= \frac{\sum_{v^{l-1}\in\cP_{l,i}}
      \Pr\{V_l \neq v^{(d,i)}, V^{l-1} = v^{l-1}\}}
    {\sum_{v^{l-1}\in\cP_{l,i}} \Pr\{V^{l-1} = v^{l-1}\}}\notag\\
    &= \sum_{v^{l-1}\in\cP_{l,i}} \frac{\Pr\{V_l \neq v^{(d,i)}| V^{l-1} = v^{l-1}\} \Pr\{V^{l-1} = v^{l-1}\} }
    {\sum_{v^{l-1}\in\cP_{l,i}} \Pr\{V^{l-1} = v^{l-1}\}}\notag\\
    &= \sum_{v^{l-1}\in\cP_{l,i}} (1-\Pr\{V_l = v^{(d,i)}| V^{l-1} = v^{l-1}\}) \notag \\
    &\quad \times \frac{ \Pr\{V^{l-1} = v^{l-1}\}}
    {\sum_{v^{l-1}\in\cP_{l,i}} \Pr\{V^{l-1} = v^{l-1}\}}\notag\\
    &\eqo{(a)} \frac{(l-1)\delta - 2 (l-2)-1}{(l-1)\delta - 2 (l-2)}
    \frac{\sum_{v^{l-1}\in\cP_{l,i}}
      \Pr\{V^{l-1} = v^{l-1}\}}
    {\sum_{v^{l-1}\in\cP_{l,i}}
      \Pr\{V^{l-1} = v^{l-1}\}}\notag\\
    &= \frac{(l-1)\delta - 2 (l-2)-1}{(l-1)\delta - 2 (l-2)},
    \label{equ:prob of first partial event Ei}
\end{align}
where $\cP_{l,i} = \{ v^{l-1}\in\cS_{l-1}: v_{j_h}=v^{(d,h)},\ \forall h \in \{1,\cdots,i-1\},
v_{m}\neq v^{(d,i)},\ \forall m\in\{j_{i-1}+1,\cdots,l-1\}\}$, and (a) comes from (\ref{equ:prob of vn given vn-1}). Similarly, we have
\begin{align}
    &\Pr\{V_{j_i} = v^{(d,i)} | \cap_{l=j_{i-1}+1}^{j_i-1}V_l \neq v^{(d,i)}, \cap_{m=1}^{i-1} \cE_m \}\notag\\
    &= \Pr\{V_{j_i} = v^{(d,i)} | \cup_{v^{j_i-1}\in\cP_i}\{V^{j_i-1}=v^{j_i-1}\}\}\notag\\
    &= \frac{\Pr\{V_{j_i} = v^{(d,i)}, \cup_{v^{j_i-1}\in\cP_i}\{V^{j_i-1}=v^{j_i-1}\}\}}
    {\Pr\{\cup_{v^{j_i-1}\in\cP_i}\{V^{j_i-1}=v^{j_i-1}\}\}}\notag\\
    &= \frac{\sum_{v^{j_i-1}\in\cP_i}\Pr\{V_{j_i} = v^{(d,i)}, V^{j_i-1}=v^{j_i-1}\}}
    {\sum_{v^{j_i-1}\in\cP_i}\Pr\{V^{j_i-1}=v^{j_i-1}\}}\notag\\
    &= \sum_{v^{j_i-1}\in\cP_i} \Pr\{V_{j_i} = v^{(d,i)}|V^{j_i-1}=v^{j_i-1}\}\notag\\
    &\quad \times \frac{\Pr\{V^{j_i-1}=v^{j_i-1}\}}
    {\sum_{v^{j_i-1}\in\cP_i}\Pr\{V^{j_i-1}=v^{j_i-1}\}}\notag\\
    &= \frac{1}{(j_i-1)\delta - 2 (j_i-2)},
    \label{equ:prob of sec partial event Ei}
\end{align}
where
\begin{align*}
    \cP_{i}\teq& \big\{v^{j_i-1}\in\cS_{j_i-1}: v_{j_h}=v^{(d,h)}\ \forall h \in \{1,\cdots,i-1\},\notag\\
    & v_{l}\neq v^{(d,i)}\ \forall l\in\{j_{i-1}+1,\cdots,j_{i}-1\}\big\}.
\end{align*}

By substituting (\ref{equ:prob of first partial event Ei}) and (\ref{equ:prob of sec partial event Ei}) into (\ref{equ:prob of event Ei}), we have \eqref{equ:from_app_a}.

\section{} \label{app:bivariate_GF_SN1}
Let $f(u, z)$ be a double series defined as
\begin{align*}
    f(u, z) \teq \sum_{k, l = 0}^\infty \frac{k - 1}{k!} s(k - 1,l) u^l z^k,
\end{align*}
where we assume that $s(-1, l) = 0$. First of all, we show that $f(u, z)$ is absolutely convergent.

If we assume that $\left[{-1 \atop l}\right] = 0$, we have
\begin{align*}     
    &\sum_{k = 0}^{\infty} \sum_{l = 0}^{\infty} \frac{k}{k!} \left[{k - 1 \atop l}\right] u^l z^k\\
    &= \sum_{k = 1}^{\infty} \sum_{l = 0}^{\infty} \frac{k}{k!} \left[{k - 1 \atop l}\right] u^l z^k\\
    &= z \sum_{k = 0}^{\infty} \sum_{l = 0}^{\infty} \frac{k + 1}{(k + 1)!} \left[{k \atop l}\right] u^l z^{k}\\
    &= z \sum_{k = 0}^{\infty} \sum_{l = 0}^{\infty} \frac{1}{k!} \left[{k \atop l}\right] u^l z^{k}\\
    &= z \sum_{k=0}^{\infty} \frac{1}{k!} \left( \sum_{l=0}^{\infty} \left[{k \atop l}\right] u^l \right) z^k\\
    &\eqo{(a)} z \sum_{k=0}^{\infty} \frac{1}{k!} u (u + 1) \cdots (u + k - 1) z^k\\
    &= z \sum_{k=0}^{\infty} \frac{1}{k!} (-u) (- u - 1) \cdots (- u - k + 1) (-1)^k z^k\\
    &= z \sum_{k=0}^{\infty} \frac{(-u) (- u - 1) \cdots (- u - k + 1)}{k!} (-z)^k\\
    &= z \sum_{k=0}^{\infty} \binom{-u}{k} (-z)^k\\
    &\eqo{(b)} z (1 - z)^{- u} \quad (\forall u \in \mathbb{R},\, \forall z \in \mathbb{R} \mbox{ s.t. } |z| < 1),
\end{align*}
where $\binom{a}{k}$ denotes the generalized binomial coefficient defined as for any real number $a \in \mathbb{R}$,
\begin{align*}
    \binom{a}{k} \teq \frac{a(a - 1) \cdots (a - k + 1)}{k!},
\end{align*}
(a) follows since
\begin{align*}
    x^{\overline{k}} = \sum_{l = 0}^{\infty} \left[{k \atop l}\right] x^l,
\end{align*}
and (b) comes from the Maclaurin series for $(1 + z)^a$ which is convergent if $|z| < 1$. Since $|z (1 - z)^{- u}| < \infty$ for any $u \in \mathbb{R}$ and $z \in \mathbb{R}$ such that $|z| < 1$, the above iterated series is convergent. According to \cite[Proposition 212]{montesinos2015introduction}, if $u \geq 0$ and $z \in [0, 1)$, the double series is also convergent, i.e.,
\begin{align*}
    \sum_{k, l = 0}^{\infty} \frac{k}{k!} \left[{k - 1 \atop l}\right] u^l z^k
    = \sum_{k=0}^{\infty} \sum_{l=0}^{\infty} \frac{k}{k!} \left[{k - 1 \atop l}\right] u^l z^k < \infty.
\end{align*}
Since for any $u, z, k, l \geq 0$,
\begin{align*}
    0 \leq \frac{k - 1}{k!} \left[{k - 1 \atop l}\right] u^l z^k \leq \frac{k}{k!} \left[{k - 1 \atop l}\right] u^l z^k,
\end{align*}
we also have, according to \cite[Corollary 210]{montesinos2015introduction},
\begin{align*}
    \sum_{k, l = 0}^{\infty} \frac{k - 1}{k!} \left[{k - 1 \atop l}\right] u^l z^k < \infty \quad (\forall u \geq 0, \forall z \in [0, 1)).
\end{align*}
Now, for any $u \in \mathbb{R}$ and $z \in \mathbb{R}$ such that $|z| < 1$, we have
\begin{align*}
    &\sum_{k, l = 0}^{\infty} \left| \frac{k - 1}{k!} s(k - 1, l) u^l z^k \right|\\
    &= \sum_{k, l = 0}^{\infty} \left| \frac{k - 1}{k!} (-1)^{k - 1 - l} \left[{k - 1 \atop l}\right] u^l z^k \right|\\
    &= \sum_{k, l = 0}^{\infty} \frac{k - 1}{k!} \left[{k - 1 \atop l}\right] |u|^l |z|^k
    < \infty.
\end{align*}
This means that $f(u, z)$ is absolutely convergent.

We note that, according to this fact and \cite[Proposition 213]{montesinos2015introduction}, iterated series are equivalent for any $u \in \mathbb{R}$ and $z \in \mathbb{R}$ such that $|z| < 1$, i.e.,
\begin{align}
    \sum_{l=0}^\infty \sum_{k = 0}^{\infty} \frac{k - 1}{k!} s(k - 1,l) z^k u^l
    = \sum_{k = 0}^{\infty} \sum_{l=0}^\infty \frac{k - 1}{k!} s(k - 1,l) u^l z^k.
    \label{equ:app:abs_conv}
\end{align}

Let 
\begin{align*}
    f_{l}(z) &\teq \sum_{k = 0}^{\infty} \frac{k - 1}{k!} s(k - 1,l) z^k.
\end{align*}
Since
\begin{align*}
    \frac{1}{z} f_{l}(z)
    &= \frac{1}{z} \sum_{k = 0}^{\infty} \frac{k - 1}{k!} s(k - 1,l) z^k\\
    &= \frac{1}{z} \sum_{k = 1}^{\infty} \frac{k - 1}{k!} s(k - 1,l) z^k\\
    &= \frac{1}{z} z \sum_{k = 0}^{\infty} \frac{k}{(k + 1)!} s(k,l) z^{k}\\
    &= \sum_{k=l}^{\infty} \frac{k}{k+1} \frac{z^k}{k!} s(k,l),
\end{align*}
we need a closed-form expression of $\frac{1}{z}f_{l}(z)$ for $z = - \frac{1}{2}$. To this end, we evaluate the following series:
\begin{align*}
    &\sum_{l=0}^\infty f_{l}(z) u^l\\
    &= \sum_{l=0}^\infty \sum_{k = 0}^{\infty} \frac{k - 1}{k!} s(k - 1,l) z^k u^l\\
    &\eqo{(a)} \sum_{k = 0}^{\infty} \sum_{l=0}^\infty \frac{k - 1}{k!} s(k - 1,l) u^l z^k\\
    &= \sum_{k = 1}^{\infty} \frac{k - 1}{k!} \left(\sum_{l=0}^\infty s(k - 1,l) u^l\right) z^k\\
    &\eqo{(b)} \sum_{k = 1}^{\infty} \frac{k - 1}{k!} u(u-1)\cdots(u-k+2) z^k \\
    &= \sum_{k = 1}^{\infty} \frac{k - 1}{u + 1} \frac{(u + 1)u(u-1)\cdots(u-k+2)}{k!} z^k \\
    &= \sum_{k=1}^{\infty} \frac{k - 1}{u + 1} \binom{u + 1}{k} z^k\\
    &\eqo{(c)} \frac{1}{u + 1} + \sum_{k=0}^{\infty} \frac{k - 1}{u + 1} \binom{u + 1}{k} z^k\\
    &= \frac{1}{u + 1} + \sum_{k=0}^{\infty} \frac{k}{u + 1} \binom{u + 1}{k} z^k\\
    &\quad - \sum_{k=0}^{\infty} \frac{1}{u + 1} \binom{u + 1}{k} z^k\\
    &= \frac{1}{u + 1} + z \sum_{k=0}^{\infty} \frac{k + 1}{u + 1} \binom{u + 1}{k + 1} z^k\\
    &\quad - \sum_{k=0}^{\infty} \frac{1}{u + 1} \binom{u + 1}{k} z^k\\
    &= \frac{1}{u + 1} + z \sum_{k=0}^{\infty} \binom{u}{k} z^k - \frac{1}{u + 1}  \sum_{k=0}^{\infty} \binom{u + 1}{k} z^k\\
    &\eqo{(d)} \frac{1}{u+1} + z (1+z)^u - \frac{1}{u+1} (1+z)^{u+1}\\
    &= \frac{1}{u+1} + z (1+z)^u - (1 + z) \frac{1}{u+1} (1+z)^{u}\\
    &\eqo{(e)} \sum_{l=0}^\infty (-1)^l u^l + z\sum_{l=0}^\infty \frac{(\ln(1+z))^l}{l!} u^l\\
    &\quad - (1 + z) \left( \sum_{l=0}^\infty (-1)^l u^l \right) \left(\sum_{l=0}^\infty  \frac{(\ln(1+z))^l}{l!} u^l \right)\\
    &= \sum_{l=0}^\infty (-1)^l u^l + z\sum_{l=0}^\infty \frac{(\ln(1+z))^l}{l!} u^l\\
    &\quad - (1 + z) \sum_{l=0}^\infty  \left(\sum_{m=0}^l \frac{(\ln(1+z))^m}{m!} (-1)^{l - m}\right)u^l\\
    &= \sum_{l=0}^\infty (-1)^l u^l
    + z\sum_{l=0}^\infty \frac{(\ln(1+z))^l}{l!} u^l\\
    &\quad - (1 + z) \sum_{l=0}^\infty  (-1)^{l}\left(\sum_{m=0}^l \frac{(-\ln(1+z))^m}{m!}\right)u^l\\
    &= \sum_{l=0}^\infty \left( z \frac{(\ln(1+z))^l}{l!} \right.\\
    &\quad  \left. + (-1)^{l} \left(1 - (1 + z)  \sum_{m=0}^l \frac{(-\ln(1+z))^m}{m!} \right) \right)u^l,
\end{align*}
where (a) comes from \eqref{equ:app:abs_conv}, (b) follows since $\sum_{l=0}^{\infty} s(k,l) u^l = u(u-1)\cdots(u-k+1)$, (c) comes from the fact that $\binom{u + 1}{k} = 1$ if $k = 0$, (d) comes from Maclaurin series with respect to $z$ which are convergent if $|z| < 1$, and (e) comes from Maclaurin series with respect to $u$ which are convergent if $|u| < 1$.

Thus, for any $z, u \in \mathbb{R}$ such that $|z| < 1$ and $|u| < 1$, we have
\begin{align*}
    \sum_{l = 0}^{\infty} f_{l}(z) u^{l}&= \sum_{l = 0}^{\infty} \left( z \frac{(\ln(1+z))^l}{l!} \right.\\
    &\quad \left. + (-1)^{l}\left(1 - (1 + z) \sum_{m=0}^l \frac{(-\ln(1+z))^m}{m!} \right) \right) u^{l}.
\end{align*}

Since two power series are convergent in a neighborhood of $0$, all coefficients are equal (see \cite[Corollary 3.8]{ fischer2012course}). This means that
\begin{align*}
    f_{l}(z) &= z \frac{(\ln(1+z))^l}{l!}\\
    &\quad + (-1)^{l}\left(1 - (1 + z) \sum_{m=0}^l \frac{(-\ln(1+z))^m}{m!} \right),
\end{align*}
where $|z| < 1$. Thus, we have
\begin{align*}
    \frac{1}{z} f_{l}(z) &=  \frac{(\ln(1+z))^l}{l!}\\
    &\quad + (-1)^{l}\left(\frac{1}{z}
        - \frac{1+z}{z}
        \sum_{m=0}^l \frac{(-\ln(1+z))^m}{m!}
    \right).
\end{align*}
Especially, when $z = -\frac{1}{2}$, we have
\begin{align*}
    &-2 f_{l}(- 1 / 2)\\
    &= \frac{(\ln(1/2))^l}{l!} + (-1)^{l}\left(- 2 + \sum_{m=0}^l \frac{(-\ln(1 / 2))^m}{m!} \right)\\
    &= (-1)^l\frac{(\ln 2)^l}{l!} + (-1)^{l}\left(- 2 + \sum_{m=0}^l \frac{( \ln 2)^m}{m!} \right)\\
    &= (-1)^{l}\left(\frac{(\ln 2)^l}{l!} - 2 + \sum_{m=0}^l \frac{( \ln 2)^m}{m!} \right).
\end{align*}

\section{} \label{app:rumor_centrality}
In this appendix, we prove Lemma \ref{lem:correct prob with cond.}.

First of all, we introduce some notations. Let $R(v, \cG_n)$ be the rumor centrality \cite{shah2011rumors} of a node $v$ in $\cG_n$, $T_{w}^{v}$ be the subtree of $\cG_n$ rooted at the node $w$ with the ancestor node $v$, and $|T_{w}^{v}|$ be the number of nodes in $T_{w}^{v}$. Here, we assume that $T_{w}^{v} = \emptyset$ and $|T_{w}^{v}| = 0$ if $w \notin \cV(\cG_n)$. We note that the ML estimator becomes (see.\ \cite[Section II-C]{shah2011rumors})
\begin{align*}
    \varphi_{\mathrm{ML}}(\cG_n) = \argmax_{v\in\cV(\cG_n)} R(v, \cG_n).
\end{align*}

Consider a sub-neighborhood $\cN_l(v)\subseteq \cN(v)$, where $\cN(v)$ is the set of neighboring nodes of $v$ in the graph $\cG$. For $v \in \cV(\cG_n)$, if $R(v,\cG_n)\geq R(w,\cG_n)$ for all $w\in \cN_{l}(v) \cap \cV(\cG_n)$, then $v$ is called the local rumor center w.r.t.\ $\cN_l(v)$. For the local rumor center, we know the following properties (see.\ \cite[Proposition 1]{dong2013rootingArXiv4}):
\begin{itemize}
    \item For a node $v \in \cV(\cG_n)$, it holds that $|T_{w}^{v}|\leq \frac{n}{2}$ for all $w\in \cN_l(v)$
    $\Leftrightarrow$ the node $v$ is a local rumor center w.r.t.\ $\cN_l(v)$.
    \item A node $v \in \cV(\cG_n)$ is a local rumor center w.r.t.\ $\cN_l(v)$ $\Rightarrow$ it holds that
    \begin{align*}
        R(w', \cG_n) < R(v, \cG_n), \quad \forall w' \in \bigcup_{w \in \cN_l(v)} \{T_{w}^{v} \setminus \{w\} \}.
    \end{align*}
    \item A node $v \in \cV(\cG_n)$ is a local rumor center w.r.t.\ $\cN_l(v)$ $\Rightarrow$ there exists at most a node $w\in \cN_l(v)$ such that       
    \begin{align*}
        R(w, \cG_n) = R(v, \cG_n),
    \end{align*}
    where the equality holds if and only if
    \begin{align*}
        |T_{w}^{v}|= \frac{n}{2}.
    \end{align*}
\end{itemize}

According to these properties, for a node $v \in \cV(\cG_n)$, if it holds that $|T_{w}^{v}|\leq \frac{n}{2}$ for all $w\in \cN(v)$, the node $v$ is a (local) rumor center w.r.t.\ $\cN(v)$. Then, there exists at most a node $w\in \cN(v)$ such that
\begin{align*}
    R(w', \cG_n)&< R(v, \cG_n), \ \forall w' \in \cV(\cG_n) \backslash \{v,w\},
\end{align*}
and
\begin{align*}
    R(w, \cG_n) = R(v, \cG_n),
\end{align*}
where the equality holds if and only if
\begin{align*}
    |T_{w}^{v}|= \frac{n}{2}.
\end{align*}
Hence, for a node $v \in \cV(\cG_n)$, if $\overline{X}(v) < n/2$, i.e., $\max\{|T_{w}^{v}|, w\in \cN(v)\}<n/2$,
we have
\begin{align*}
    R(w', \cG_n)&< R(v, \cG_n), \ \forall w' \in \cV(\cG_n) \backslash \{v\}.
\end{align*}
Thus, the MAP estimator outputs $v$, and hence
\begin{align*}
    \Pr\{\hat V_{n}=v| v \in \cV_n, \overline X(v) < n/2 \}&= 1.
\end{align*}

For a node $v \in \cV(\cG_n)$, if $\overline{X}(v) =\frac{n}{2}$, i.e., there exists a node $w\in \cN(v)$ such that $|T_{w}^v|=\frac{n}{2}$ and $|T_{w'}^v|<\frac{n}{2}$ for any other $w'\in \cN(v)$, we have
\begin{align*}
    R(w', \cG_n)&< R(v, \cG_n), \ \forall w' \in \cV(\cG_n) \backslash \{v,w\},
\end{align*}
and
\begin{align*}
    R(w, \cG_n)= R(v, \cG_n).
\end{align*}
Thus, the MAP estimator outputs $v$ with probability $1 / 2$, and hence
\begin{align*}
    \Pr\{\hat V_{n}=v| v \in \cV_n, \overline X(v) = n/2 \}&= 1/2.
\end{align*}

For a node $v \in \cV(\cG_n)$, if $\overline{X}(v) > n/2$, i.e., $\max\{|T_{w}^{v}|, w\in \cN(v)\}>n/2$, the node $v$ is not a local rumor center w.r.t.\ $\cN(v)$. Hence there exists $w\in \cN(v)$ such that
\begin{align*}
    R(w, \cG_n)> R(v, \cG_n).
\end{align*}
Then, the MAP estimator does not output $v$, and hence
\begin{align*}
    \Pr\{\hat V_{n}=v| v \in \cV_n, \overline X(v) > n/2 \}&= 0.
\end{align*}
This completes the proof.

\section{} \label{app:zeta_function}
We note that
\begin{align*}
    \zeta_{k}^{d}(0) = \sum_{1\leq j_1 <j_2<\cdots j_d \leq k} \frac{1}{j_1j_2\cdots j_d}.
\end{align*}
and
\begin{align*}
    \zeta_{k}^{d}(1) = \sum_{1\leq j_1 < j_2<\cdots<j_{d} \leq k} \frac{1}{(j_1+1)\cdots (j_d+1)}.
\end{align*}
Thus, for any $d \geq 1$ and $k \geq d$, we have
\begin{align*}
    &\zeta_{k}^{d}(1)\\
    &= \sum_{j_1=1}^{k-d+1}\sum_{j_2=j_1+1}^{k-d+2}
    \cdots \sum_{j_i=j_{i-1}+1}^{k-d+i}
    \cdots \sum_{j_d=j_{d-1}+1}^{k-d+d}\\
    &\quad \times \frac{1}{(j_1+1)\cdots (j_d+1)}\\
    &= \sum_{j_1=1}^{k-d+1}\sum_{j_2=j_1+1}^{k-d+2}
    \cdots \sum_{j_i=j_{i-1}+1}^{k-d+i}
    \cdots \sum_{j_d=j_{d-1}+1}^{k-d+d}\\
    &\quad \times \frac{1}{(j_1+1)\cdots (j_d+1)}\\
    &\quad + \sum_{j_1=0}^{0}\sum_{j_2=j_1+1}^{k-d+2}
    \cdots \sum_{j_i=j_{i-1}+1}^{k-d+i}
    \cdots \sum_{j_d=j_{d-1}+1}^{k-d+d}\\
    &\quad \times \frac{1}{(j_1+1)\cdots (j_d+1)}\\
    &\quad - \sum_{j_1=0}^{0}\sum_{j_2=j_1+1}^{k-d+2}
    \cdots \sum_{j_i=j_{i-1}+1}^{k-d+i}
    \cdots \sum_{j_d=j_{d-1}+1}^{k-d+d}\\
    &\quad \times \frac{1}{(j_1+1)\cdots (j_d+1)}\\
    &= \sum_{j_1=0}^{k-d+1}\sum_{j_2=j_1+1}^{k-d+2}
    \cdots \sum_{j_i=j_{i-1}+1}^{k-d+i}
    \cdots \sum_{j_d=j_{d-1}+1}^{k-d+d}\\
    &\quad \times \frac{1}{(j_1+1)\cdots (j_d+1)}\\
    &\quad - \sum_{j_2=1}^{k-d+2}
    \cdots \sum_{j_i=j_{i-1}+1}^{k-d+i}
    \cdots \sum_{j_d=j_{d-1}+1}^{k-d+d}\\
    &\quad \times \frac{1}{(j_2+1)\cdots (j_d+1)}\\
    &= \sum_{j_1=1}^{k+1-d+1}\sum_{j_2=j_1}^{k-d+2}\sum_{j_3=j_2 + 1}^{k-d+3}
    \cdots \sum_{j_i=j_{i-1}+1}^{k-d+i}
    \cdots \sum_{j_d=j_{d-1}+1}^{k-d+d}\\
    &\quad \times \frac{1}{j_1(j_2 + 1)\cdots (j_d+1)}\\
    &\quad - \sum_{j_2=1}^{k-d+2}
    \cdots \sum_{j_i=j_{i-1}+1}^{k-d+i}
    \cdots \sum_{j_d=j_{d-1}+1}^{k-d+d}\\
    &\quad \times \frac{1}{(j_2+1)\cdots (j_d+1)}\\
    &= \sum_{j_1=1}^{k+1-d+1}\sum_{j_2=j_1+1}^{k+1-d+2} \sum_{j_3=j_2}^{k-d+3}
    \cdots \sum_{j_i=j_{i-1}+1}^{k-d+i}
    \cdots \sum_{j_d=j_{d-1}+1}^{k-d+d}\\
    &\quad \times \frac{1}{j_1j_2(j_3 + 1)\cdots (j_d+1)}\\
    &\quad - \sum_{1\leq j_2<j_3<\cdots<j_d\leq k}
    \frac{1}{(j_2+1)\cdots (j_d+1)}\\
    &= \sum_{j_1=1}^{k+1-d+1}\sum_{j_2=j_1+1}^{k+1-d+2}
    \cdots \sum_{j_i=j_{i-1}+1}^{k+1-d+i}
    \cdots \sum_{j_d=j_{d-1}+1}^{k+1-d+d}
    \frac{1}{j_1j_2\cdots j_d}\\
    &\quad - \sum_{1\leq j_2<j_3<\cdots<j_d\leq k}
    \frac{1}{(j_2+1)\cdots (j_d+1)}\\
    &= \zeta_{k+1}^{d}(0)
    - \sum_{1\leq j_1<j_2<\cdots<j_{d-1}\leq k}
    \frac{1}{(j_1+1)\cdots (j_{d-1}+1)}\\
    &=  \zeta_{k+1}^{d}(0) - \zeta_{k}^{d-1}(1),
\end{align*}
where $\zeta_{k}^{0}(1) = 1$.

\section{} \label{app:limit_of_Dn}
In order to show the equation \eqref{equ:limit_of_Dn}, we use the next lemma (cf.\ e.g.\ \cite{tao2011introduction}).
\begin{lem}[Dominated convergence theorem]
    \label{lem:dct}
    Let $f_1,f_2,\cdots: \mathbb{N} \to \mathbb{R}$ be a sequence of real-valued functions on positive integers $\mathbb{N}$ such that
    \begin{align*}
        f_{n}(k) &\mbox{ converges as $n \to \infty$}, \quad \forall k \in \mathbb{N}.
    \end{align*}
    Suppose that there is $g: \mathbb{N} \to \mathbb{R}$ such that
    \begin{align*}
        \sum_{k = 1}^{\infty} g(k) &< \infty,\\
        |f_{n}(k)| &\leq g(k), \quad \forall n, k \in \mathbb{N}.
    \end{align*}
    Then, we have
    \begin{align*}
        \lim_{n \to \infty} \sum_{k = 1}^{\infty} f_{n}(k) = \sum_{k = 1}^{\infty} \lim_{n \to \infty} f_{n}(k).
    \end{align*}
\end{lem}

We note that
\begin{align*}
    &\Pr\{v^{(d)}\in\cV_n, \overline X(v^{(d)})<n/2\}\\
    &= \sum_{k=d+1}^{\lceil n/2 \rceil}\Pr\{V_k=v^{(d)}, \overline X(v^{(d)})<n/2\}\\
    &\eqo{(a)} \sum_{k=1}^{\lceil n/2 \rceil}\Pr\{V_k=v^{(d)}, \overline X(v^{(d)})<n/2\}\\
    &\eqo{(b)} \sum_{k=1}^{\infty}\Pr\{V_k=v^{(d)}, \overline X(v^{(d)})<n/2\},
\end{align*}
where (a) follows since $\Pr\{V_k=v^{(d)}\} = 0$ for any $k \leq d$, and (b) comes from the fact that if $v^{(d)}$ is the $k$th infected node ($k \geq \lceil n/2 \rceil + 1$), it must hold that $\overline X(v^{(d)}) \geq n/2$. We also note that
\begin{align*}
    &\Pr\{v^{(d)}\in\cV_n, \overline X(v^{(d)})=n/2\}\\
    &= \sum_{k=d+1}^{\lfloor n/2 \rfloor + 1}\Pr\{V_k=v^{(d)}, \overline X(v^{(d)}) = n/2\}\\
    &= \sum_{k=1}^{\lfloor n/2 \rfloor + 1}\Pr\{V_k=v^{(d)}, \overline X(v^{(d)}) = n/2\}\\
    &\eqo{(a)} \sum_{k=1}^{\infty}\Pr\{V_k=v^{(d)}, \overline X(v^{(d)}) = n/2\},
\end{align*}
where (a) comes from the fact that if $v^{(d)}$ is the $k$th infected node ($k \geq \lfloor n/2 \rfloor + 2$), it must hold that $X(v^{(d)}) > n/2$. Thus, we have
\begin{align*}
    \bfD_{n}(d)
    &= \sum_{v^{(d)} \in \cV^{(d)}} \left(\Pr\{v^{(d)}\in\cV_n, \overline X(v^{(d)})<n/2\}\right.\notag\\
    &\quad \left. +1/2\Pr\{v^{(d)}\in\cV_n, \overline X(v^{(d)})=n/2\}\right)\\
    &= \sum_{v^{(d)} \in \cV^{(d)}} \sum_{k=1}^{\infty} \left(\Pr\{V_k=v^{(d)}, \overline X(v^{(d)}) < n/2\}\right.\notag\\
    &\quad \left. +1/2 \Pr\{V_k=v^{(d)}, \overline X(v^{(d)}) = n/2\}\right),    
\end{align*}

By noticing that $\Pr\{V_k=v^{(d)}\}$ does not depend on $n$ (see \eqref{equ:prob vk}), we can set
\begin{align*}
    f_{n}(k) &= \Pr\{V_k=v^{(d)}, \overline X(v^{(d)})<n/2\} \\
    &\quad + 1/2 \Pr\{V_k=v^{(d)}, \overline X(v^{(d)}) = n/2\},\\
    g(k) &= \Pr\{V_k=v^{(d)}\}.
\end{align*}
Then, we have
\begin{align*}
    |f_{n}(k)| &=\Pr\{V_k=v^{(d)}, \overline X(v^{(d)})<n/2\} \\
    &\quad + 1/2 \Pr\{V_k=v^{(d)}, \overline X(v^{(d)}) = n/2\}\\
    &\leq \Pr\{V_k=v^{(d)}\} \left(\Pr\{\overline X(v^{(d)})<n/2 | V_k=v^{(d)}\}\right. \\
    &\quad \left. + 1/2 \Pr\{\overline X(v^{(d)}) = n/2 | V_k=v^{(d)}\}\right)\\
    &\leq \Pr\{V_k=v^{(d)}\} \Pr\{\overline X(v^{(d)}) \leq n/2 | V_k=v^{(d)}\}\\
    &\leq g(k).
\end{align*}
We also have
\begin{align*}
    \sum_{k = 1}^{\infty} g(k) &= \sum_{k = 1}^{\infty} \Pr\{V_k=v^{(d)}\}\\
    &= \Pr\left\{ \bigcup_{k = 1}^{\infty} \{V_k = v^{(d)} \} \right\}\\
    &\leq 1.
\end{align*}
On the other hand, according to \eqref{equ:D_for_odd_with_str_num} and \eqref{equ:D_for_even_with_str_num}, we have for any $k \geq d + 1$ and odd $n \geq 3$,
\begin{align*}
    f_{n}(k) = \frac{2}{k+1} 
    \frac{\binom{(n+3)/2}{k+1}}
    {\binom{n+1}{k+1}}
    \frac{(-1)^{d+k}}{(k-1)!}\sum_{l=1}^{d} s(k,l),
\end{align*}
and for any $k \geq d + 1$ and even $n \geq 2$,
\begin{align*}
    f_{n}(k) = \frac{2}{k+1} 
    \frac{\binom{n/2+1}{k+1}
      +  \frac{n}{2(n+2)}
      \binom{n/2+1}{k}}{\binom{n+1}{k+1}} \frac{(-1)^{d+k}}{(k-1)!}\sum_{l=1}^{d} s(k,l).
\end{align*}
By noticing that
\begin{align*}
    \lim_{m\ra\infty}\frac{\binom{((2 m + 1) + 3)/2}{k+1} }{\binom{(2 m + 1)+1}{k+1}}
    &= \lim_{m\ra\infty}\frac{\binom{m + 2}{k+1} }{\binom{2 m + 2}{k+1}}\\
    &= \frac{1}{2^{k+1}}
\end{align*}
and
\begin{align*}
    &\lim_{m\ra\infty} \frac{\binom{(2 m)/2+1}{k+1} + \frac{(2 m)}{2((2 m)+2)} \binom{(2 m)/2+1}{k}}{\binom{(2 m)+1}{k+1}}\\
    &\lim_{m\ra\infty} \frac{\binom{m +1}{k+1} + \frac{m}{2(m+1)} \binom{m+1}{k}}{\binom{2 m+1}{k+1}}\\
    &= \frac{1}{2^{k+1}},
\end{align*}
we have for any $k \geq d + 1$,
\begin{align*}
    \lim_{n \ra \infty} f_{n}(k) = \frac{1}{2^{k}} \frac{1}{k+1} \frac{ (-1)^{d+k}}{(k-1)!}\sum_{l=1}^{d} s(k,l).
\end{align*}
We note that for any $k \leq d$,
\begin{align*}
    \lim_{n \ra \infty} f_{n}(k) = 0.
\end{align*}

Thus, according to Lemma \ref{lem:dct}, we have
\begin{align}
    &\lim_{n \ra \infty} \bfD_{n}(d)\notag\\
    &= \lim_{n \ra \infty} \sum_{v^{(d)} \in \cV^{(d)}} \sum_{k=1}^{\infty} \left(\Pr\{V_k=v^{(d)}, \overline X(v^{(d)}) < n/2\}\right.\notag\notag\\
    &\quad \left. +1/2 \Pr\{V_k=v^{(d)}, \overline X(v^{(d)}) = n/2\}\right)\notag\\
    &= \lim_{n \ra \infty} \sum_{v^{(d)} \in \cV^{(d)}} \sum_{k=1}^{\infty} f_{n}(k)\notag\\
    &= \sum_{v^{(d)} \in \cV^{(d)}} \lim_{n \ra \infty} \sum_{k=1}^{\infty} f_{n}(k)\notag\\
    &= \sum_{v^{(d)} \in \cV^{(d)}} \sum_{k=1}^{\infty} \lim_{n \ra \infty} f_{n}(k)\notag\\
    &= \sum_{v^{(d)} \in \cV^{(d)}} \sum_{k=d+1}^{\infty} \lim_{n \ra \infty} f_{n}(k)\notag\\
    &= \sum_{v^{(d)} \in \cV^{(d)}} \sum_{k=d+1}^{\infty} \frac{1}{2^{k}} \frac{1}{k+1} \frac{ (-1)^{d+k}}{(k-1)!}\sum_{l=1}^{d} s(k,l).   
    \label{equ:app:limit_of_Dn}
\end{align}
By noticing that $|\cV^{(d)}| = \delta (\delta - 1)^{d - 1}$, we have \eqref{equ:limit_of_Dn} from \eqref{equ:app:limit_of_Dn}.

\section{} \label{app:polya}
We consider the P\'olya's urn model with $2$ colors balls: Initially, $b_j$ balls of color $C_j$ $(j=1,2)$ are in the urn. At each step, a single ball is uniformly drawn form the urn. Then, the drawn ball is returned with additional $m$ balls of the same color. Repeat this drawing process $n$ times. Let $\tilde Y_j$ denote the number of balls of the color $C_j$ in the urn at the end of time $n$. Let $Y_j$ denote the number of times that the balls of color $C_j$ are drawn after $n$ draws.

According to \cite[Theorem 4.1]{shah2016finding}, we have the next theorem.
\begin{thm}
    \label{thm:polya_conv_beta}
    \begin{align*}
        \frac{\tilde Y_{1}}{b_1 + b_2 + n \cdot m} \xrightarrow{\mathrm{a.s.}} Y\ (n \to \infty),
    \end{align*}
    where $b_1 + b_2 + n \cdot m$ is the total number of balls in the urn at the end of time $n$, and $Y$ is a Beta random variable with parameters $b_1/m$ and $b_2/m$. That is for $x \in [0, 1]$,
    \begin{align*}
        \Pr\{Y \leq x\} = I_{x}\left(\frac{b_1}{m}, \frac{b_2}{m}\right).
    \end{align*}    
\end{thm}

We immediately have the next corollary.
\begin{cor}
    \label{cor:polya_conv_beta}
    \begin{align*}
        \frac{Y_{1}}{n} \xrightarrow{\mathrm{a.s.}} Y,
    \end{align*}
    where $Y$ is the same Beta random variable as that of Theorem
    \ref{thm:polya_conv_beta}.
\end{cor}
\begin{IEEEproof}
    $Y_1$ can be written as
    \begin{align*}
        Y_{1} = \frac{\tilde Y_{1} - b_{1}}{m}.
    \end{align*}
    Thus, we have
    \begin{align*}
        \frac{Y_{1}}{n} &= \frac{\tilde Y_{1} - b_{1}}{m \cdot n}\\
        &= \frac{\tilde Y_{1}}{m \cdot n} - \frac{b_{1}}{m \cdot n}\\
        &= \frac{\tilde Y_{1}}{b_1 + b_2 + n \cdot m}\frac{b_1 + b_2 + n \cdot m}{m \cdot n} - \frac{b_{1}}{m \cdot n}.
    \end{align*}
    Since $\frac{b_1 + b_2 + n \cdot m}{m \cdot n} \to 1$ and $\frac{b_{1}}{m \cdot n} \to 0$ as $n \to \infty$, we have
    \begin{align*}
        \frac{Y_{1}}{n}
        &= \frac{\tilde Y_{1}}{b_1 + b_2 + n \cdot m}\frac{b_1 + b_2 + n \cdot m}{m \cdot n} - \frac{b_{1}}{m \cdot n}\\
        &\xrightarrow{\mathrm{a.s.}} Y,
    \end{align*}
    where almost sure convergence comes from Theorem \ref{thm:polya_conv_beta}. This completes the proof.
\end{IEEEproof}

After $v^{(d)}$ is infected $k$th, $X_1(v^{(d)})$ can be regarded as the P\'olya's urn model with the following settings: $Y_{1} = X_1(v^{(d)})$, $Y_{2} = \sum_{j = 2}^{\delta} X_{j}(v^{(d)}) - k + 1$, $b_{1} = 1$, $b_{2} = (k - 1)(\delta - 2) + \delta - 1$, and $m = \delta - 2$. Here, we assume that the total number of drawing balls is $n-k$. Then, according to Corollary \ref{cor:polya_conv_beta}, we have
\begin{align*}
    \frac{X_1(v^{(d)})}{n} = \frac{Y_1}{n} = \frac{Y_1}{n - k}\frac{n - k}{n} \xrightarrow{\mathrm{a.s.}} Y,
\end{align*}
where $Y$ is a Beta random variable with parameters $1/(\delta - 2)$ and $k - 1 + (\delta - 1)/(\delta - 2)$. Thus, we have
\begin{align}
    & \lim_{n\ra\infty}\Pr\{\cE_{1}(v^{(d)}) | V_k=v^{(d)}\}\notag\\
    &= \lim_{n\ra\infty}\Pr\{X_1(v^{(d)}) < n/2 | V_k=v^{(d)}\}\notag\\
    &= \lim_{n\ra\infty}\Pr\{X_1(v^{(d)})/n < 1/2 | V_k=v^{(d)}\}\notag\\
    &= \Pr\{Y < 1/2 \}\notag\\
    &= I_{1/2}\left(\frac{1}{\delta-2}, k-1+\frac{\delta-1}{\delta-2}\right)\notag\\
    &= 1 - I_{1/2}\left(k-1+\frac{\delta-1}{\delta-2}, \frac{1}{\delta-2}\right).
    \label{equ:E1_is_beta}
\end{align}
Similarly, we have
\begin{align}
    & \lim_{n\ra\infty}\Pr\{\cF_{1}(v^{(d)}) | V_k=v^{(d)}\}\notag\\
    &= \lim_{n\ra\infty}\Pr\{X_1(v^{(d)}) \leq n/2 | V_k=v^{(d)}\}\notag\\
    &= \Pr\{Y \leq 1/2 \}\notag\\
    &= 1 - I_{1/2}\left(k-1+\frac{\delta-1}{\delta-2}, \frac{1}{\delta-2}\right).
    \label{equ:F1_is_beta}
\end{align}
Due to \eqref{equ:E1_is_beta} and \eqref{equ:F1_is_beta}, we have \eqref{equ:F1=E1=I}.

On the other hand, after $v^{(d)}$ is infected $k$th, $X_\delta(v^{(d)})$ can be regarded as the P\'olya's urn model with the following settings: $Y_{1} = X_\delta(v^{(d)}) - k + 1$, $Y_{2} = \sum_{j = 1}^{\delta - 1} X_{j}(v^{(d)})$, $b_{1} = (k - 1)(\delta - 2) + 1$, $b_{2} = \delta - 1$, and $m = \delta - 2$. Here, we assume that the total number of drawing balls is $n-k$. Then, according to Corollary \ref{cor:polya_conv_beta}, we have
\begin{align*}
    \frac{X_\delta(v^{(d)})}{n} = \frac{Y_1 + k - 1}{n} = \frac{Y_1 + k - 1}{n - k}\frac{n - k}{n} \xrightarrow{\mathrm{a.s.}} Y,
\end{align*}
where $Y$ is a Beta random variable with parameters $k - 1 + 1/(\delta - 2)$ and $(\delta - 1) / (\delta - 2)$. Thus, we have
\begin{align}
    & \lim_{n\ra\infty}\Pr\{\cE_{\delta}(v^{(d)}) | V_k=v^{(d)}\}\notag\\
    &= \lim_{n\ra\infty}\Pr\{X_{\delta}(v^{(d)}) < n/2 | V_k=v^{(d)}\}\notag\\
    &= \lim_{n\ra\infty}\Pr\{X_{\delta}(v^{(d)})/n < 1/2 | V_k=v^{(d)}\}\notag\\
    &= \Pr\{Y < 1/2 \}\notag\\
    &= I_{1/2}\left(k-1+\frac{1}{\delta-2}, \frac{\delta - 1}{\delta-2}\right).
    \label{equ:Ed_is_beta}
\end{align}
Similarly, we have
\begin{align}
    & \lim_{n\ra\infty}\Pr\{\cF_{\delta}(v^{(d)}) | V_k=v^{(d)}\}\notag\\
    &= \lim_{n\ra\infty}\Pr\{X_{\delta}(v^{(d)}) \leq n/2 | V_k=v^{(d)}\}\notag\\
    &= \Pr\{Y \leq 1/2 \}\notag\\
    &= I_{1/2}\left(k-1+\frac{1}{\delta-2}, \frac{\delta - 1}{\delta-2}\right).
    \label{equ:Fd_is_beta}
\end{align}
Due to \eqref{equ:Ed_is_beta} and \eqref{equ:Fd_is_beta}, we have \eqref{equ:Fd=Ed=I}.

\section{} \label{app:E=F}
According to (\ref{equ:upper bound of dist prob using E}) and (\ref{equ:upper bound of dist prob using F}), it holds that
\begin{align*}
    &\bfD_{n}(d)\\
    &= \sum_{v^{(d)} \in \cV^{(d)}} \Pr\{v^{(d)}\in\cV_n, \hat V_{n} = v^{(d)} \}\notag\\
    &\geq \sum_{v^{(d)} \in \cV^{(d)}} \sum_{k=d+1}^{\lceil n/2 \rceil} \Pr\{V_k=v^{(d)}\} \left( \Pr\{ \cE_{\delta}(v^{(d)}) | V_k=v^{(d)}\} \right. \notag\\
    &\quad \left. - (\delta-1) (1-\Pr\{\cE_{1}(v^{(d)}) | V_k=v^{(d)}\}) \right),
\end{align*}
and
\begin{align*}
    &\bfD_{n}(d)\\
    &= \sum_{v^{(d)} \in \cV^{(d)}} \Pr\{v^{(d)}\in\cV_n, \hat V_{n} = v^{(d)} \}\notag\\
    &\leq \sum_{v^{(d)} \in \cV^{(d)}} \sum_{k=d+1}^{\lfloor n/2 \rfloor + 1} \Pr\{V_k=v^{(d)}\} \left( \Pr\{ \cF_{\delta}(v^{(d)}) | V_k=v^{(d)}\} \right. \notag\\
    &\quad \left. - (\delta-1) (1-\Pr\{\cF_{1}(v^{(d)}) | V_k=v^{(d)}\}) \right).
\end{align*}

For $k \leq \lceil n/ 2 \rceil$, we set
\begin{align*}
    f_{n}(k) &= \Pr\{V_k=v^{(d)}\} \left( \Pr\{ \cE_{\delta}(v^{(d)}) | V_k=v^{(d)}\} \right. \notag\\
    &\quad \left. - (\delta-1) (1-\Pr\{\cE_{1}(v^{(d)}) | V_k=v^{(d)}\}) \right).
\end{align*}
For $k \geq \lceil n/ 2 \rceil + 1$, we set $f_{n}(k) = 0$. According to \eqref{equ:F1=E1=I} and \eqref{equ:Fd=Ed=I}, we have for any $k \geq d + 1$,
\begin{align*}
    \lim_{n \to \infty} f_{n}(k)  = p_{1}(\delta, d, k) ( I^{(2)}(\delta, k) - (\delta-1)I^{(1)}(\delta, k) ),
\end{align*}
and for any $k \leq d$,
\begin{align*}
    \lim_{n \to \infty} f_{n}(k)  = 0.
\end{align*}
On the other hand, we set
\begin{align*}
    g(k) &= \delta \Pr\{V_k=v^{(d)}\}.
\end{align*}

Obviously, for $k \geq \lceil n/ 2 \rceil + 1$, it holds that $|f_{n}(k)| \leq g(k)$. For $k \leq \lceil n/ 2 \rceil$, we have
\begin{align*}
    |f_{n}(k)| &=\Pr\{V_k=v^{(d)}\} \left| \Pr\{ \cE_{\delta}(v^{(d)}) | V_k=v^{(d)}\} \right. \notag\\
    &\quad \left. - (\delta-1) (1-\Pr\{\cE_{1}(v^{(d)}) | V_k=v^{(d)}\}) \right|\\
    &\leq \Pr\{V_k=v^{(d)}\} \left(\left| \Pr\{ \cE_{\delta}(v^{(d)}) | V_k=v^{(d)}\} \right| \right. \notag\\
    &\quad \left. + \left| (\delta-1) (1-\Pr\{\cE_{1}(v^{(d)}) | V_k=v^{(d)}\}) \right| \right)\\
    &\leq \delta \Pr\{V_k=v^{(d)}\}\\
    &= g(k).
\end{align*}
We also have
\begin{align*}
    \sum_{k = 1}^{\infty} g(k) &= \delta \sum_{k = 1}^{\infty} \Pr\{V_k=v^{(d)}\}\\
    &= \delta \Pr\left\{ \bigcup_{k = 1}^{\infty} \{V_k = v^{(d)} \} \right\}\\
    &\leq \delta.
\end{align*}

Thus, according to Lemma \ref{lem:dct}, we have
\begin{align}
    &\liminf_{n \to \infty} \bfD_{n}(d)\notag\\
    &\geq \liminf_{n \to \infty} \sum_{v^{(d)} \in \cV^{(d)}} \sum_{k=d+1}^{\lceil n/2 \rceil} \Pr\{V_k=v^{(d)}\}\notag\\
    &\quad \times \left( \Pr\{ \cE_{\delta}(v^{(d)}) | V_k=v^{(d)}\} \right. \notag\notag\\
    &\quad \left. - (\delta-1) (1-\Pr\{\cE_{1}(v^{(d)}) | V_k=v^{(d)}\}) \right)\notag\\
    &= \liminf_{n \to \infty} \sum_{v^{(d)} \in \cV^{(d)}} \sum_{k=1}^{\lceil n/2 \rceil} \Pr\{V_k=v^{(d)}\}\notag\\
    &\quad \times \left( \Pr\{ \cE_{\delta}(v^{(d)}) | V_k=v^{(d)}\} \right. \notag\notag\\
    &\quad \left. - (\delta-1) (1-\Pr\{\cE_{1}(v^{(d)}) | V_k=v^{(d)}\}) \right)\notag\\
    &\geq \sum_{v^{(d)} \in \cV^{(d)}} \liminf_{n \to \infty} \sum_{k=1}^{\lceil n/2 \rceil} \Pr\{V_k=v^{(d)}\}\notag\\
    &\quad \times \left( \Pr\{ \cE_{\delta}(v^{(d)}) | V_k=v^{(d)}\} \right. \notag\notag\\
    &\quad \left. - (\delta-1) (1-\Pr\{\cE_{1}(v^{(d)}) | V_k=v^{(d)}\}) \right)\notag\\
    &= \sum_{v^{(d)} \in \cV^{(d)}} \liminf_{n \to \infty} \sum_{k=1}^{\lceil n/2 \rceil} f_{n}(k)\notag\\
    &= \sum_{v^{(d)} \in \cV^{(d)}} \liminf_{n \to \infty} \sum_{k=1}^{\infty} f_{n}(k)\notag\\
    &= \sum_{v^{(d)} \in \cV^{(d)}} \sum_{k=1}^{\infty} \lim_{n \to \infty} f_{n}(k)\notag\\
    &= \sum_{v^{(d)} \in \cV^{(d)}} \sum_{k=d+1}^{\infty} \lim_{n \to \infty} f_{n}(k)\notag\\    
    &= \sum_{v^{(d)} \in \cV^{(d)}} \sum_{k=d+1}^{\infty} p_{1}(\delta, d, k) ( I^{(2)}(\delta, k) - (\delta-1)I^{(1)}(\delta, k) ). \label{equ:D_leq_ank}
\end{align}

For $k \leq \lfloor n/ 2 \rfloor + 1$, we set
\begin{align*}
    h_{n}(k) &= \Pr\{V_k=v^{(d)}\} \left( \Pr\{ \cF_{\delta}(v^{(d)}) | V_k=v^{(d)}\} \right. \notag\\
    &\quad \left. - (\delta-1) (1-\Pr\{\cF_{1}(v^{(d)}) | V_k=v^{(d)}\}) \right).
\end{align*}
For $k \geq \lfloor n/ 2 \rfloor + 2$, we set $h_{n}(k) = 0$. According to \eqref{equ:F1=E1=I} and \eqref{equ:Fd=Ed=I}, we have for any $k \geq d + 1$, 
\begin{align*}
    \lim_{n \to \infty} h_{n}(k)  = p_{1}(\delta, d, k) ( I^{(2)}(\delta, k) - (\delta-1)I^{(1)}(\delta, k) ),
\end{align*}
and for any $k \leq d$,
\begin{align*}
    \lim_{n \to \infty} h_{n}(k) = 0.
\end{align*}
We also have $|h_{n}(k)| \leq g(k)$.

Thus, according to Lemma \ref{lem:dct}, we have
\begin{align}
    &\limsup_{n \to \infty} \bfD_{n}(d)\notag\\
    &\leq \limsup_{n \to \infty} \sum_{v^{(d)} \in \cV^{(d)}} \sum_{k=d+1}^{\lfloor n/ 2 \rfloor + 1} \Pr\{V_k=v^{(d)}\}\notag\\
    &\quad \times \left( \Pr\{ \cF_{\delta}(v^{(d)}) | V_k=v^{(d)}\} \right. \notag\notag\\
    &\quad \left. - (\delta-1) (1-\Pr\{\cF_{1}(v^{(d)}) | V_k=v^{(d)}\}) \right)\notag\\
    &\leq \sum_{v^{(d)} \in \cV^{(d)}} \limsup_{n \to \infty} \sum_{k=1}^{\lfloor n/ 2 \rfloor + 1} \Pr\{V_k=v^{(d)}\}\notag\\
    &\quad \times \left( \Pr\{ \cF_{\delta}(v^{(d)}) | V_k=v^{(d)}\} \right. \notag\notag\\
    &\quad \left. - (\delta-1) (1-\Pr\{\cF_{1}(v^{(d)}) | V_k=v^{(d)}\}) \right)\notag\\
    &= \sum_{v^{(d)} \in \cV^{(d)}} \limsup_{n \to \infty} \sum_{k=1}^{\lfloor n/ 2 \rfloor + 1} h_{n}(k)\notag\\
    &= \sum_{v^{(d)} \in \cV^{(d)}} \limsup_{n \to \infty} \sum_{k=1}^{\infty} h_{n}(k)\notag\\
    &= \sum_{v^{(d)} \in \cV^{(d)}} \sum_{k=1}^{\infty} \lim_{n \to \infty} h_{n}(k)\notag\\
    &= \sum_{v^{(d)} \in \cV^{(d)}} \sum_{k=d+1}^{\infty} \lim_{n \to \infty} h_{n}(k)\notag\\
    &= \sum_{v^{(d)} \in \cV^{(d)}} \sum_{k=d+1}^{\infty} p_{1}(\delta, d, k) ( I^{(2)}(\delta, k) - (\delta-1)I^{(1)}(\delta, k) ).
    \label{equ:D_geq_cnk}
\end{align}

By noticing that $|\cV^{(d)}| = \delta (\delta - 1)^{d - 1}$, we have \eqref{equ:dist prob with lim for any degree} from \eqref{equ:D_leq_ank} and \eqref{equ:D_geq_cnk}.



\begin{thebibliography}{10}
    \providecommand{\url}[1]{#1}
    \csname url@samestyle\endcsname
    \providecommand{\newblock}{\relax}
    \providecommand{\bibinfo}[2]{#2}
    \providecommand{\BIBentrySTDinterwordspacing}{\spaceskip=0pt\relax}
    \providecommand{\BIBentryALTinterwordstretchfactor}{4}
    \providecommand{\BIBentryALTinterwordspacing}{\spaceskip=\fontdimen2\font plus
      \BIBentryALTinterwordstretchfactor\fontdimen3\font minus
      \fontdimen4\font\relax}
    \providecommand{\BIBforeignlanguage}[2]{{%
        \expandafter\ifx\csname l@#1\endcsname\relax
        \typeout{** WARNING: IEEEtran.bst: No hyphenation pattern has been}%
        \typeout{** loaded for the language `#1'. Using the pattern for}%
        \typeout{** the default language instead.}%
        \else
        \language=\csname l@#1\endcsname
        \fi
        #2}}
    \providecommand{\BIBdecl}{\relax}
    \BIBdecl

    \bibitem{matsuta2014pdd}
    T.~Matsuta and T.~Uyematsu, ``Probability distributions of the distance between
    the rumor source and its estimation on regular trees,'' in \emph{Proc. 37th
      Symp. on Inf. Theory and its Apps.}, Dec. 2014, pp. 605--610.

    \bibitem{bailey1975mathematical}
    N.~T.~J. Bailey, \emph{The Mathematical Theory of Infectious Diseases and Its
      Applications}.\hskip 1em plus 0.5em minus 0.4em\relax Charles Griffin \&
    Company Ltd., 1975.

    \bibitem{shah2011rumors}
    D.~Shah and T.~Zaman, ``Rumors in a network: Who's the culprit?'' \emph{IEEE
      Trans. Inf. Theory}, vol.~57, no.~8, pp. 5163--5181, Aug. 2011.

    \bibitem{pastor2001epidemic}
    R.~Pastor-Satorras and A.~Vespignani, ``Epidemic spreading in scale-free
    networks,'' \emph{Phys. Rev. Lett.}, vol.~86, pp. 3200--3203, Apr. 2001.

    \bibitem{may2001infection}
    R.~M. May and A.~L. Lloyd, ``Infection dynamics on scale-free networks,''
    \emph{Phys. Rev. E}, vol.~64, p. 066112, Nov. 2001.

    \bibitem{7740106}
    J.~Khim and P.~Loh, ``Confidence sets for the source of a diffusion in regular
    trees,'' \emph{IEEE Trans. on Netw. Sci. Eng.}, vol.~4, no.~1, pp. 27--40,
    Jan 2017.

    \bibitem{shah2016finding}
    D.~Shah and T.~Zaman, ``Finding rumor sources on random trees,''
    \emph{Operations Research}, vol.~64, no.~3, pp. 736--755, Feb. 2016.

    \bibitem{murty2006multiple}
    M.~R. Murty and K.~Sinha, ``{Multiple Hurwitz zeta functions},'' \emph{Proc.
      Symp. in Pure Math.}, vol.~75, pp. 135--156, 2006.

    \bibitem{hoffman1992multiple}
    M.~Hoffman, ``Multiple harmonic series,'' \emph{Pacific Journal of Math.}, vol.
    152, no.~2, pp. 275--290, Feb. 1992.

    \bibitem{dong2013rootingArXiv4}
    W.~Dong, W.~Zhang, and C.~W. Tan, ``Rooting out the rumor culprit from
    suspects,'' \emph{Arxiv preprint arXiv:1301.6312v4}, May. 2013.

    \bibitem{johnson1977urn}
    N.~L. Johnson and S.~Kotz, \emph{{Urn Models and Their Application: An Approach
        to Modern Discrete Probability Theory}}.\hskip 1em plus 0.5em minus
    0.4em\relax Wiley New York, 1977.

    \bibitem{adamchik1997stirling}
    V.~Adamchik, ``{On Stirling numbers and Euler sums},'' \emph{Journal of
      Computational and Applied Math.}, vol.~79, no.~1, pp. 119--130, 1997.

    \bibitem{kuba2010note}
    M.~Kuba and H.~Prodinger, ``{A note on Stirling series},'' \emph{Integers},
    vol.~10, no.~4, pp. 393--406, 2010.

    \bibitem{graham1994concrete}
    R.~L. Graham, D.~E. Knuth, and O.~Patashnik, \emph{Concrete Mathematics: A
      Foundation for Computer Science}.\hskip 1em plus 0.5em minus 0.4em\relax
    Addison-Wesley Longman Publishing Co., Inc., 1994.

    \bibitem{montesinos2015introduction}
    V.~Montesinos, P.~Zizler, and V.~Zizler, \emph{An introduction to modern
      analysis}.\hskip 1em plus 0.5em minus 0.4em\relax Springer, 2015.

    \bibitem{fischer2012course}
    W.~Fischer and I.~Lieb, \emph{A Course in Complex Analysis}.\hskip 1em plus
    0.5em minus 0.4em\relax Springer, 2012.

    \bibitem{tao2011introduction}
    T.~Tao, \emph{An Introduction to Measure Theory}.\hskip 1em plus 0.5em minus
    0.4em\relax American Mathematical Soc., 2011, vol. 126.

\end{thebibliography}



\end{document}